\documentclass[11pt, a4paper]{article}
\pdfoutput=1
\usepackage{jcappub}

\usepackage[utf8]{inputenc}

\usepackage{lipsum}

\usepackage{color}

\usepackage{graphicx} 
\usepackage{float}
\usepackage{subcaption}  
\usepackage{hyperref}   

\usepackage{fancyhdr} 

\usepackage{mathtools}
\usepackage{amssymb}
\usepackage{cancel}
\usepackage{units}
\usepackage{bigints}
\usepackage{array}
\numberwithin{equation}{section}

\DeclareMathOperator{\tr}{tr}
\DeclareMathOperator{\cst}{const}

\begin{document}

\newcommand{\rg}{\sqrt{-g}}
\newcommand{\rf}{\sqrt{-f}}
\newcommand{\ovl}[1]{\overline{#1}}
\newcommand{\mn}{^{\mu}_{\ \ \nu}}
\newcommand{\mmn}{^{(m)\mu}_{\ \ \ \ \ \ \nu}}
\newcommand{\cH}{\mathcal{H}}
\newcommand{\wt}[1]{\widetilde{#1}}
\def\lsim{\raisebox{-.4ex}{$\stackrel{<}{\scriptstyle \sim}$\,}}
\def\gsim{\raisebox{-.4ex}{$\stackrel{>}{\scriptstyle \sim}$\,}}
\title{Cosmology with moving bimetric fluids}

\author{Carlos García-García, Antonio L.~Maroto, Prado Mart\'{\i}n-Moruno}
\affiliation{Departamento de F\'{\i}sica Te\'orica I, Universidad Complutense de Madrid, E-28040 Madrid, Spain.} 

\emailAdd{cargar08@ucm.es}
\emailAdd{maroto@ucm.es}
\emailAdd{pradomm@ucm.es}

\abstract{We study cosmological  implications of bigravity and massive gravity
  solutions with non-simultaneously diagonal metrics by considering the
  generalized Gordon and Kerr-Schild ansatzes. The scenario that we obtain is
  equivalent to that of General Relativity with additional non-comoving perfect
  fluids. We show that the most general ghost-free bimetric theory generates
  three kinds of effective fluids whose equations of state are fixed by  a
  function of the ansatz. Different choices of such function allow to reproduce
  the behaviour of different dark fluids. In particular, the Gordon ansatz is
  suitable for the description of various kinds of slowly-moving fluids,
  whereas the Kerr-Schild one is shown to describe a null dark energy
  component.  The motion of those dark fluids with respect to the CMB is shown
  to generate, in turn, a relative motion of baryonic matter with respect to
  radition which contributes to the CMB anisotropies. CMB dipole observations
  are able to set stringent limits on the dark sector described by the
  effective bimetric fluid.  } 
  
\maketitle

\section{Introduction}
Despite the recent advances in cosmology, there are fundamental questions about
the universe that still remain unsolved. One of such questions is its
composition. Recent measurements from type Ia supernovae confirmed by cosmic
microwave background (CMB) and large scale structure observations have shown
that our Universe is expanding in an accelerated way
\cite{Perlmutter:1998np,Riess:1998cb,Ade:2015xua}. In order to explain the
observed behavior, cosmologists postulated the existence of a cosmological
constant or dark energy fluid that can only be detected by its gravitational
effects. Assuming the validity of General Relativity (GR), it should comprise
around the 68\% of the total energy density of the universe. In addition,
around the 27\% of the total content is taken to be cold dark matter, some kind
of fluid with low pressure that seems to weakly interact with ordinary matter
so that, to date, its existence can only be inferred from  its gravitational
effects. This leaves us with only a 5\% of known origin that would be composed
of baryonic matter and a negiglible part (0.01\%) of radiation. In addition,
yet another dark fluid could exist: dark radiation. However, it would only be a
10\%, at most, of the radiation energy density \cite{Ade:2015xua}.  
On the other hand, the CMB dipole, quadrupole and octupole exhibit an
unexpected allignment \cite{Liu:2013wfa,Copi:2006tu}, which might suggest the
possible existence of a preferred direction in the Universe. Furthermore, the
possible existence of a dark flow \cite{Ade:2015xua,Atrio-Barandela:2014nda},
that is, a coherent motion of matter with respect to the CMB on cosmological
scales, would also support this idea.

One could think that this complicated picture may be just a signal of the
breakdown of GR on very large scales. Following this line of thought we will
explore the cosmological implications of well-known theories as massive gravity
and bigravity.  Massive gravity dates from 1939, when Fierz and Pauli tried to
give mass to the carrier particle  of the gravitational interaction, i.e. the
graviton \cite{Fierz:1939ix}.  In the first place, it was found that the
massless limit of massive gravity could not recover GR, where the graviton only
has 2 propagating degrees of freedom in contrast to a massive graviton which
necessarily propagates 5.  This fact lies under the appearance of the vDVZ (van
Dam--Veltman--Zakharov) discontinuity \cite{vanDam:1970vg,Zakharov:1970cc}.
However, Vainshtein showed some years later that the extra degree of freedom
responsible for that discontinuity could be screened by its own interaction,
which dominates over the linear terms in the massless case
\cite{Vainshtein:1972sx}.  In the second place, non-linear massive gravity was
thought to be always affected by the Boulware--Deser instability
\cite{Boulware:1973my}, that is a state with negative kinetic energy leading to
an unbounded Hamiltonian. Nonetheless, it has been recently  discovered that
there are particular theories \cite{deRham:2010kj} in which this ghost can be avoided  \cite{Hassan:2011hr} (see also
references \cite{deRham:2014zqa,Schmidt-May:2015vnx} and references therein).  On the other hand,
the existing astrophysical and  cosmological observations can constraint the
value of the graviton mass to be $m<7.2\times \unit[10^{-23}]{eV}$. In any
case, a graviton with $m \ll \unit[10^{-33}]{eV}$ could not be distinguished
from a massless graviton \cite{Goldhaber:2008xy,deRham:2016nuf}.

Non-linear massive gravity is formulated using a non-dynamical reference metric
in addition to the dynamical one describing our spacetime \cite{Visser:1997hd}.
This bimetric nature of massive gravity connects it directly with bigravity
\cite{Isham:1971gm}, in which both metrics are dynamical, although it must be
kept in mind that they are conceptually and phenomenologically different
theories \cite{Baccetti:2012bk}.  
However, the consideration of the same term of interaction between the metrics that avoids the Boulware--Deser ghost in massive gravity \cite{deRham:2010kj} has led to the formulation of a stable bigravity theory \cite{Hassan:2011zd}.
The existence of a second gravitational
sector in bigravity allows the presence of two kinds  of matter fields, each
one minimally coupled to its respective dynamical metric.  Furthermore, the
consequences of more general kinds of couplings, which we will not consider in
the present work since they re-introduce the instability, have been also investigated 
\cite{Akrami:2013ffa,deRham:2014naa}.

Cosmological solutions of both types of ghost-free theories have been obtained
recently in the literature.  In the case of stable massive gravity theories formulated
using different reference metrics \cite{Hassan:2011vm,Hassan:2011tf}, it has been shown that either there is no
homogeneous and isotropic accelerating solutions
\cite{D'Amico:2011jj,MartinMoruno:2013gq} or they are affected by the Higuchi
instability \cite{Fasiello:2012rw}. On the other hand, although
self-accelerating cosmological solutions have been found in stable bigravity
\cite{vonStrauss:2011mq,Konnig:2013gxa}, early-time fast growing modes
\cite{Lagos:2014lca,Konnig:2014xva,Cusin:2014psa} have to be carefully avoided
\cite{Amendola:2015tua,Akrami:2015qga}. These works considered solutions where both metrics
can be written in a diagonal way in the same coordinate patch  (see reference
\cite{Nersisyan:2015oha} and references therein). In the present paper we will
go beyond that assumption studying the cosmological consequences of massive
gravity and bigravity solutions where both metrics are not necessarily diagonal
in the same coordinate patch. In order to simplify the treatment of these
bimetric theories, we will  focus our attention on solutions with particular
causal relations between both metrics, described by the generalized Gordon
ansatz and the Kerr-Schild ansatz \cite{Baccetti:2012ge}.  As we will show, the
non-diagonality of both metrics in the same coordinate patch can be
re-interpreted from the point of view of standard GR as an effective perfect
fluid which is non-comoving with respect to the CMB.  The relative motion of
the effective bimetric fluid induces, in turn, a relative motion of baryons
with respect to the CMB after recombination thus contributing to the CMB
dipole. This effect opens the possibility to  observationally constrain the
possible values of the free parameters of these moving bimetric theories.

This work is organized as follows. In section \ref{S:notes}, ghost-free massive
gravity and bigravity will be briefly reviewed. In section \ref{S:Gordon}, we
will impose the generalized Gordon ansatz and study the resulting effective
fluids in section \ref{S:Gordon:SS:split}. Its cosmological implications will
be studied in section \ref{S:FLRW}, considering the slow-moving regime
compatible with a FLRW solution.  In section \ref{S:FLRW:SS:Monocomponent}, the
monocomponent effective fluid and its effect on the  CMB dipole will be
investigated, while in section \ref{S:FLRW:SS:Multicomponent} the
multicomponent case will be considered. In section \ref{S:Kerr}, the
Kerr-Schild anstaz will be used to consider the fast-moving regime. In section
\ref{S:Kerr:SS:split}, the resulting null-fluid and its natural subcomponents
will be studied.  The cosmological case with a Bianchi I metric will be
investigated in section \ref{S:Bianchi}. Finally, in section \ref{S:Conclusion}
the conclusions will be summarized.

\section{Massive gravity and bigravity}\label{S:notes}
The action of massive gravity and that of bigravity are written in terms of two
metric tensors $g_{\mu\nu}$ and $f_{\mu\nu}$, where $g_{\mu\nu}$ is the
standard metric to which the observed physical fields are coupled, whereas
$f_{\mu\nu}$ is the new reference metric introduced with the theory in massive
gravity and an additional dynamical field in bigravity. Thus the action can be
written in general as
 \begin{align}\nonumber
    S=&-\frac{1}{16\pi G}\int d^4x \rg [R(g) + 2\Lambda - 2m^2 L_{\rm int}(g,f)] +
        S_{(m)}\\
      &- \frac{\kappa}{16 \pi G}\int d^4 x \rf [\ovl{R}(f) + 2\ovl{\Lambda}]
    + \epsilon \bar{S}_{(m)}
    \label{S:notes:eq:action}
  \end{align}
where $R$ is the curvature scalar of $g_{\mu\nu}$, overlined variables are
defined using $f_{\mu\nu}$, $m$ is the graviton mass, $G$ the gravitational
constant, and $\kappa$ and $\epsilon$ are constants which vanish in the case of
massive gravity (since $f_{\mu\nu}$ is non-dynamical). We will work in natural
units $\hbar=c=1$. We consider that the matter sectors described through
$S_{(m)}$ and $\bar{S}_{(m)}$ are minimally coupled to $g_{\mu\nu}$ and
$f_{\mu\nu}$, respectively.
These theories are ghost-free \cite{Hassan:2011hr,Hassan:2011zd} if the
interaction term is given in terms of the matrix $\gamma = \sqrt{g^{-1}f}$,
that is
\begin{equation}
  \gamma^\mu_{~\nu}\gamma^\nu_{~\rho} = g^{\mu\nu}f_{\nu\rho},
  \label{S:notes:eq:gamma}
\end{equation}
as \cite{deRham:2010kj}
\begin{equation}
  L_{\rm int} = \beta_1 e_1(\gamma) + \beta_2 e_2(\gamma) + \beta_3 e_3(\gamma),
\end{equation}
where the elementary symmetric polinomials are 
\begin{align}
  e_1(\gamma)& = \tr(\gamma),\\
  e_2(\gamma)& = \frac{1}{2} \left( \tr(\gamma)^2 - \tr(\gamma^2) \right),\\
  e_3(\gamma)& = \frac{1}{6} \left( \tr(\gamma)^3 - 3 \tr(\gamma) \tr(\gamma^2)
    + 2 \tr(\gamma^3) \right), 
\end{align}
with $\beta_n$ being arbitrary constants. It is important to note that there
are two additional terms proportional to $\beta_0$ and $\beta_4$ respectively,
that have been absorbed into the two cosmological constants $\Lambda$ and
$\ovl{\Lambda}$ \cite{Baccetti:2012bk,Baccetti:2012re} as they appear in the
interaction Lagrangian through $\beta_0 e_0 (\gamma) =\beta_0$ and $\beta_4 e_4
(\gamma)= \beta_4 \det (\gamma)$.

Varying the action (\ref{S:notes:eq:action}) with respect to the dynamical
metrics (only one for massive gravity), one obtains the Einstein equations for
both sectors \cite{Baccetti:2012bk}
\begin{equation}
  G\mn - \Lambda \delta\mn = m^2 T\mn + 8\pi G T\mmn,
  \label{S:notes:eq:Einstein-g}
\end{equation}
and
\begin{equation}
  \kappa (\ovl{G}\mn - \ovl{\Lambda} \delta\mn) = m^2 \ovl{T}\mn + \epsilon
  8\pi G \ovl{T}\mmn,
  \label{S:notes:eq:Einstein-f}
\end{equation}
where $T\mmn$ and $\ovl{T}\mmn$ are the stress-energy tensors of the material
components of each sector, while $T\mn$ and $\ovl T\mn$ are the effective
stress-energy tensors that encapsulate the interaction between both metrics.
These effective tensors are \cite{Volkov:2011an}
\begin{equation}
  T\mn = \tau \mn - \delta\mn L_{int},
\end{equation}
and
\begin{equation}
  \ovl T\mn = - \frac{\sqrt{-g}}{\sqrt{-f}}\tau\mn,
  \label{S:notes:eq:overlined-T}
\end{equation}
with
\begin{equation}
  \tau\mn = \gamma^\mu_{~\rho} \frac{\partial L_{int}}{\partial
    \gamma^\nu_{~\rho}}.
\end{equation}
It should be noted that indices in equations (\ref{S:notes:eq:Einstein-g}) and
(\ref{S:notes:eq:Einstein-f}) are raised and lowered with $g$ and $f$,
respectively. In addition, it is important to remark that in massive gravity,
equation (\ref{S:notes:eq:Einstein-f}) is absent.
As the Bianchi constrain is satisfied, one has
\begin{equation}
  \nabla_\mu T\mn = 0 \qquad\mbox{ and }\qquad \ovl{\nabla}_\mu \ovl{T}\mn = 0.
  \label{S:notes:eq:Bianchi}
\end{equation}
It can be seen that both constraints are equivalent \cite{Volkov:2011an}. Thus,
the conservation of the effective fluid in the $g$-sector implies the
conservation of the other effective fluid in the $f$-sector, and vice versa.
Finally, it is important to remark that both $T\mn$ and $\ovl T\mn$ are linear in
$\beta_i$. Then, we can redefine them so that $m^2 T\mn \rightarrow 8\pi G
T\mn$, and similarly for the overlined tensor. This way, $\beta_i \rightarrow
\beta_i' = m^2/(8\pi G) \beta_i$. We will drop the prime and use these
$\beta_i'$ in the subsequent sections.

\section{Perfect bimetric fluid}\label{S:Gordon}
In this section we will restrict our attention to a particular set of
solutions of massive gravity and bigravity. We will consider a particular relation between
the causal structures of both metrics that leads to an effective
stress-energy tensor with the form of a perfect fluid. This relation
is given by the generalized Gordon ansatz \cite{Baccetti:2012ge}:
\begin{equation}
  f_{\mu\nu} = \Omega ^2 (g_{\mu\nu} + \xi V_\mu V_\nu),
  \label{S:Gordon:eq:Gordon}
\end{equation}
where $\Omega$ and $\xi$ are arbitrary functions, and $V^\mu$ is a timelike
vector with $g_{\mu \nu} V^\mu V^\nu= -1$.  One can also work with the function
$\zeta\neq0$, defined by $\xi = 1 - \zeta^2$, that better parametrizes the fact
that we need $\xi<1$ to have both metrics with Lorentzian signature, implying
also that $V^\mu$ is timelike with respect to $f_{\mu\nu}$.
The generalized Gordon ansatz can be interpreted as a conformal transformation combined with a stretch along the timelike direction parallel to $V^\mu$.
This ansatz relates the position of the light cones of both metrics, that is the light cones of $f_{\mu\nu}$ lie strictly outside (inside) the light cones of  $g_{\mu\nu}$  if $\xi<0$ ($\xi>0$), having the metrics the same null vectors if $\xi=0$. 
It must be emphasized that we are only restricting attention to a particular kind of solutions. In the case of bigravity we are assuming that the material content in the $f$-sector is such that it generates a metric $f_{\mu\nu}$ of the form given by equation (\ref{S:Gordon:eq:Gordon}) through equations (\ref{S:notes:eq:Einstein-f}). Then, one should obtain $g_{\mu\nu}$ (completely fixing $f_{\mu\nu}$) by considering equations (\ref{S:notes:eq:Einstein-g}).
For massive gravity our approach consists in considering only theories with a reference metric of the form expressed in the ansatz (\ref{S:Gordon:eq:Gordon}). This implies that not all stable massive gravity theories that we could construct will have this kind of solutions, as $g_{\mu\nu}$ has to be determined through equation (\ref{S:notes:eq:Einstein-g}) taking into account this  $f_{\mu\nu}$ and the material content. Indeed, if one completely fixes the functions appearing in the ansatz  (\ref{S:Gordon:eq:Gordon}), then only a given kind of material content will be compatible with these solutions.
This issue will be further clarified in section \ref{S:FLRW}.

The Gordon ansatz implies that the square root matrix takes the form
\begin{equation}
  \gamma\mn = \Omega \left\{ (\delta\mn + V^\mu V_\nu) - \zeta
V^\mu V_\nu \right\},
\end{equation}
where the principal square root of the matrix, corresponding to a positive
value of $\zeta$, was taken in reference \cite{Baccetti:2012ge}. Therefore, the
stress-energy tensor of the effective fluid takes the perfect fluid form
\cite{Baccetti:2012ge} \begin{equation} T\mn = (\rho_{BF} + p_{BF}) V^\mu V_\nu
  + p_{BF} \delta\mn, \label{S:Gordon:eq:T} \end{equation}
where BF stands for bimetric fluid and with
\begin{align}
  \rho_{BF} =& \Omega(3 \beta_1 + 3 \beta_2 \Omega + \beta_3 \Omega^2),
  \label{S:Gordon:eq:density}\\
  p_{BF} =& -\Omega \left[ (2\beta_1 + \beta_2\Omega) + (\beta_1 + 2\beta_2 \Omega +
  \beta_3 \Omega^2)\zeta  \right].
  \label{S:Gordon:eq:pressure}
\end{align}
Furthermore, one can obtain that the effective stress-energy tensor for the
$f$-sector is \cite{Baccetti:2012ge}
\begin{equation}
  \ovl T\mn = (\ovl \rho_{BF} + \ovl p_{BF}) \ovl V^\mu \ovl V_\nu + \ovl p_{BF} \delta\mn,
  \label{S:Gordon:eq:overlined-T}
\end{equation}
with
\begin{align}
  \ovl \rho_{BF} =& - \frac{1}{\Omega^3} (\beta_1 + 3\beta_2 \Omega + 3\beta_3 \Omega^2)  
  \label{S:Gordon:eq:overlined-hat-rho}\\
  \ovl p_{BF} =&
  -\frac{1}{\Omega^3 \zeta}\left[ \beta_1 + 3\beta_2 \Omega(2+\zeta) + \beta_3 \Omega^2
  (1+2\zeta)\right]
  \label{S:Gordon:eq:overlined-hat-p}
\end{align}
and $\ovl V_\mu = \Omega \zeta V_\mu$ with $f_{\mu\nu} \ovl V^\mu\ovl V^\nu =
-1$.  Note that the pre-factors in equations
(\ref{S:Gordon:eq:overlined-hat-rho}) and (\ref{S:Gordon:eq:overlined-hat-p})
are due to the term $\sqrt{-f} = \Omega^4 \zeta \sqrt{-g}$ in equation
(\ref{S:notes:eq:overlined-T}). In the present paper we will not restrict our
study to the principal square root, $\zeta>0$, due to the phenomenological
interest of solutions with $\zeta<0$. It should be pointed out, however, that
when considering the branch $\zeta<0$ one should consistently take all the
negative signed square roots in the $f$-sector, in particular those appearing
in the  $f$-volume element $d^4x \sqrt{-f}$ and in the vector $\ovl V^\mu$,
which would be pointing to the opposite direction to $V^\mu$. 
Consequently, solutions of the negative branch may propagate a spin-2 ghost.
Therefore, the stability and consistency of solutions with $\zeta<0$ should be carefully
considered.

\subsection{Interacting effective perfect fluids}\label{S:Gordon:SS:split}
As we have summarized, in massive gravity and bigravity the effects of the
interaction of both metrics can be encapsulated in an effective stress-energy
tensor, which takes the form of a perfect fluid when one restrict attention to
solutions satisfying the generalized Gordon ansatz. We can go further and split
the effective fluid in other three different perfect fluids that will share a
common velocity and evolve in a way that conserve energy and momentum as a
whole, but not necessarily separately. Taking into account equations
(\ref{S:Gordon:eq:density}), (\ref{S:Gordon:eq:pressure}),
(\ref{S:Gordon:eq:overlined-hat-rho}), and
(\ref{S:Gordon:eq:overlined-hat-p}), such a natural decomposition yields
\begin{align}\nonumber
  \rho_1 &=  3\beta_1 \Omega, &p_1 &=  -\beta_1 \Omega (2 + \zeta),
   &\omega_1 &=  -\frac{1}{3} (2+\zeta);\\ \nonumber
  \rho_2 &=  3\beta_2 \Omega ^2, &p_2 &= - \beta_2 \Omega^2 (1 + 2\zeta),
  &\omega_2 &=  -\frac{1}{3} (1+2\zeta);\\
  \rho_3 &=  \beta_3 \Omega^3, &p_3 &=  -\beta_3 \Omega^3 \zeta,
  &\omega_3 &=  -\zeta; 
  \label{S:Gordon:eq:fluid3}
\end{align}
and
\begin{align}\nonumber
  \ovl\rho_1 &=  \beta_1 \Omega^{-3}, &\ovl p_1 &=  -\beta_1
    \Omega^{-3} \zeta^{-1}, &\ovl\omega_1 &=  -\zeta^{-1};\\\nonumber
  \ovl\rho_2 &=  3\beta_2 \Omega^{-2},  &\ovl p_2 &= - \beta_2
    \Omega^{-2} \frac{2 + \zeta}{\zeta}, &\ovl\omega_2 &=
    -\frac{2+\zeta}{3\zeta};\\
  \ovl\rho_3 &=  3\beta_3 \Omega^{-1}, &\ovl p_3 &=  -\beta_3
    \Omega^{-1} \frac{1+2\zeta}{\zeta}, &\ovl\omega_3 &=
    -\frac{1+2\zeta}{3\zeta}.
\end{align}
The fluids of both sectors have the same effective equation of state parameter
when $\omega_n=\ovl\omega_n =\left\{ -1, 1/3 \right\}$. It must be noted that
the first case corresponds to that of two conformally related metrics, which is
known to be equivalent to having an extra contribution to the cosmological
constant for both sectors.  Moreover, both effective densities can be expressed
in general by
\begin{equation}
  \rho_n = \rho_{n0} \left( \Omega/\Omega_0 \right)^n \mbox{ and }~ 
  \ovl \rho_n = \ovl\rho_{n0} \left( \Omega/\Omega_0 \right)^{n-4},
\end{equation}
with $n=1,2,3$; and $\ovl \omega_n$ can be written in terms of $\omega_n$ as
\begin{equation}
  \ovl\omega_1 = \frac{1}{3\omega_1 +2}, ~~ \ovl\omega_2 =
  \frac{1-\omega_2}{3\omega_2 + 1}, ~~ \ovl\omega_3 =
  \frac{1-2\omega_3}{3\omega_3}.
\end{equation}

\begin{figure}[htb]
  \begin{subfigure}[b]{0.49\textwidth}
  \centering
  \includegraphics[width=\textwidth]{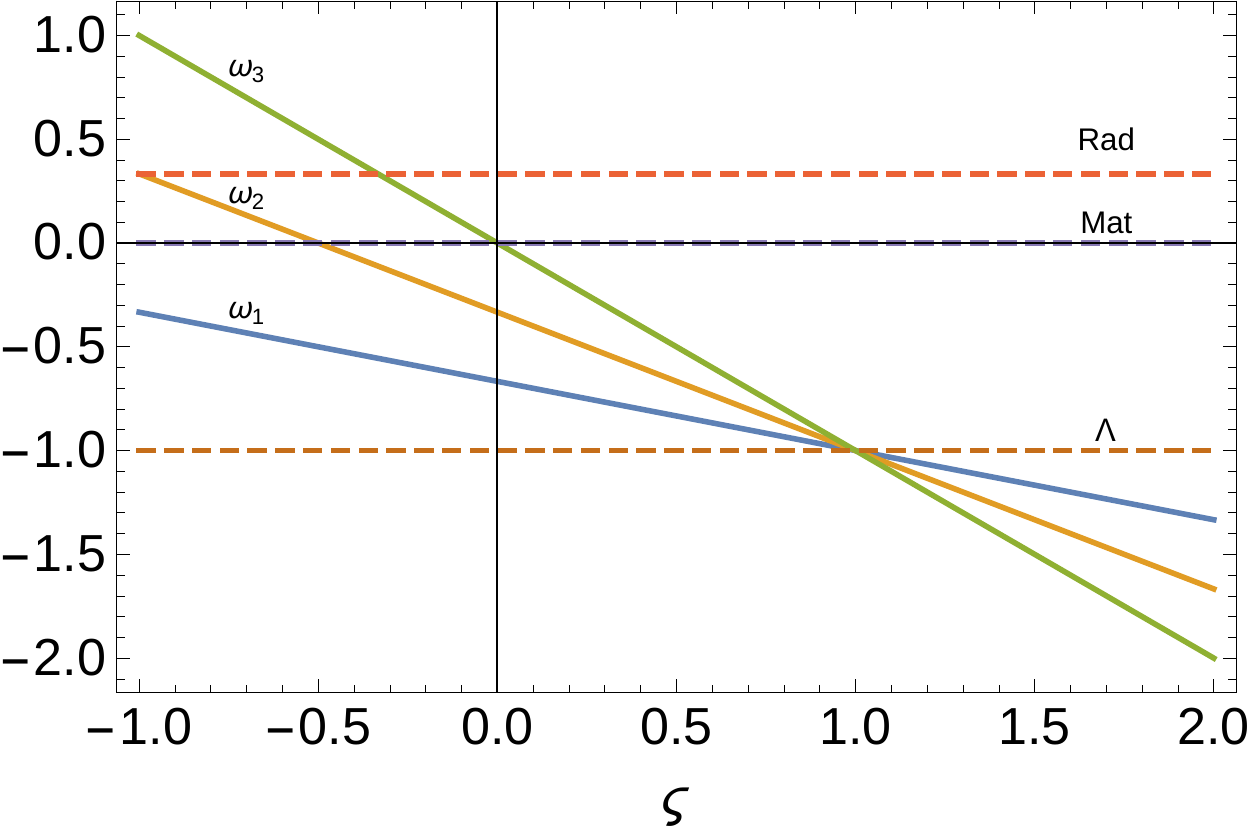}
  \end{subfigure}
  \begin{subfigure}[b]{0.49\textwidth}
    \includegraphics[width=\textwidth]{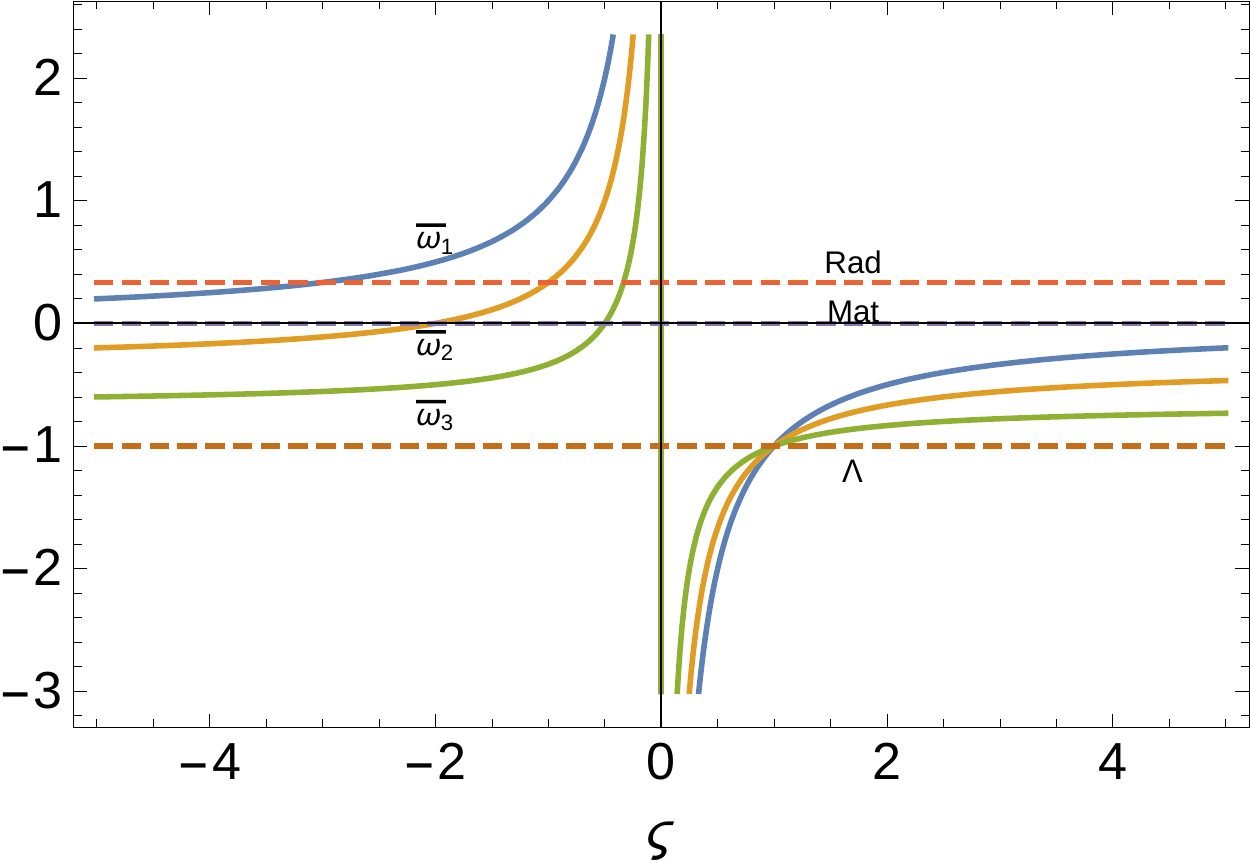}
  \end{subfigure}
  \caption{Equation of state parameter of the effective fluid's components of the $g$-sector (left) and the $f$-sector (right) as a function of $\zeta$. As it should be expected, the case $\zeta=0$ is not well defined.} 
    \label{S:Gordon:fig:plotW-z}
\end{figure}

In order to better understand the nature of these effective fluids,
$\omega(\zeta)$ and $\ovl \omega(\zeta)$  have been plotted in figure
\ref{S:Gordon:fig:plotW-z}, where we have restricted to values $\omega_n
\lesssim1$. As one could expect from the arguments at the beginning of this
section, the $\ovl\omega$ plot reveals that the case $\zeta=0$ is not well
defined.  It must be noted that for $\zeta = 1$, both metrics are conformally
related and $\omega_n = \ovl\omega_n= -1$ for $n=1,2,3$. 
Finally, it should be emphasized that two more effective components are generated
by these bimetric theories when $\beta_0 \neq 0$ and $\beta_4 \neq 0$, which, 
as mentioned before, 
naturally describe a cosmological constant in each gravitational sector.

\section{Slow-moving bimetric fluid: FLRW solutions}\label{S:FLRW}

In this section we will study cosmological solutions based on the generalized
Gordon ansatz introduced above. Thus, we will restrict ourselves to the case in
which the free functions appearing in the ansatz are just function of time,
i.e.  $\Omega=\Omega(\eta)$, $\xi=\xi(\eta)$ and $V^\mu = a^{-1} ( 1, \vec
v_{BF}(\eta))$, where we have assumed $|\vec v_{BF}|\ll 1$
in order to avoid the generation of large anisotropies. Then, the
stress-energy tensor for each fluid will be of the form
\begin{equation}
  T^\mu_{(\alpha)\nu} =  (\rho_{(\alpha)} + p_{(\alpha)}) V_{(\alpha)}^\mu
  V_{(\alpha)\,\nu} + p_{(\alpha)} \delta\mn,
  \label{S:FLRW:eq:T}
\end{equation}
with $\rho_{(\alpha)}$ the density function, $p_{(\alpha)}$ the pressure, and
$V_{(\alpha)}^\mu = a^{-1} ( 1, \vec{v}_{(\alpha)}(\eta))$ the velocity for the
fluid $\alpha = R, DR, DM, B, BF,\cdots$ (radiation, dark radiation, dark
matter, baryonic matter, bimetric fluid, \ldots) which we  also assume to be
small, i.e.  $\vert \vec v_{(\alpha)}\vert\ll 1$. We will use $\alpha$ for all
fluids in our universe, including the bimetric one, while we will use the
subscript $n=1,2,3$ for the bimetric fluid subcomponents.  Therefore, a metric
compatible with this scenario in the $g$-sector is
\begin{equation}
  ds_g^2 = a^2 (\eta) \left(-d\eta^2 - 2S_i
    d\eta dx^i + (\delta_{ij} + 2F_{ij}) dx^{i}dx^{j}\right),
    \label{S:FLRW:eq:FLRW-perturbed}
\end{equation}
where $S_i$ and $F_{ij}$ are two homogeneous time-dependent functions.

To first order in velocities, the total stress-energy tensor is \cite{Maroto:2005kc}
\begin{align}
  T^0_{~0} =& - \sum_\alpha \rho_{(\alpha)}, \\
  T^i_{~0} =& -\sum_\alpha (\rho_{(\alpha)} + p_{(\alpha)}) v_{(\alpha)}^i,\\
  T^0_{~i} =& \sum_\alpha (\rho_{(\alpha)} + p_{(\alpha)}) (v_{(\alpha)}^i-S^i),\\
  T^i_{~j} =& \sum_\alpha p_{(\alpha)} \delta^i_j,
\end{align}
where the sum goes through the different kind of fluids.  Considering the
modified Einstein equations (\ref{S:notes:eq:Einstein-g}), we see from the
$(^i_{\;j})$ component that $F_{ij}$ vanishes to first order in velocities and
from the  $(^0_{\;i})$ components, we get
\begin{equation}
  \vec S = \dfrac{\sum_\alpha (\rho_\alpha +
    p_\alpha) \vec v_{\alpha}}{\sum_\alpha (\rho_\alpha + p_\alpha)},
  \label{S:FLRW:eq:S}
\end{equation}
revealing that $\vec S$ is the cosmic center of mass velocity
\cite{Maroto:2005kc}. We can choose to work in the center of mass frame with
$\vec S = \vec 0$, leaving the metric as
\begin{equation}
  ds^2_g = a(\eta)^2 [-d\eta^2 + \delta_{ij} dx^i dx^j].
  \label{S:FLRW:eq:g}
\end{equation}
Considering equation (\ref{S:Gordon:eq:Gordon}), the metric in the $f$-sector is
\begin{equation}
  ds^2_f = \Omega^2 a^2 \left[ (-1 + \xi) d\eta^2 - 2 \xi \vec v_{BF} d\eta d\vec x + 
  \delta_{ij} dx^i dx^j \right]
  \label{S:FLRW:eq:f}
\end{equation}
It can be checked that it is indeed Lorentzian for all $\xi<1$. It is easy to
note that when $V^\mu$ is the comoving timelike vector, both metrics will be
FLRW and expressed in a diagonal way in the same coordinate patch, as it was
seen in reference \cite{Baccetti:2012ge}. The fact that the ansatz entails a stretching along the timelike direction parallel to $V^\mu$, which is not necessarily parallel to the comoving timelike vector,
is the ultimate
responsible for having a non diagonal $f_{\mu\nu}$ and a moving bimetric dark
fluid with respect to the comoving coordinates. Finally, let us note that the
stress energy tensor for the material components of the $f$-sector, described
by $\ovl T\mmn$, is also of the perfect fluid form. It can be seen from the
modified Einstein equations (\ref{S:notes:eq:Einstein-f}) that if there were no
material components in the $f$-sector, $\ovl G\mn \propto \ovl T\mn$, where
$\ovl T\mn$ is known from equation (\ref{S:Gordon:eq:overlined-T}) to be a
perfect fluid, as well. Then, $\ovl G\mn$ must be compatible with perfect
fluids solutions. As a consequence, if both $\ovl T\mn$ and $\ovl G\mn$ are to
preserve this character, the additional component $\ovl T\mmn$ must be a
perfect fluid as well.

We can now find the conservation equations in the $g$-sector up to first order
in $v_\alpha$, which are the same as those obtained in general relativity with
moving fluids \cite{Maroto:2005kc}. These are:
\begin{itemize}
  \item Energy conservation
    \begin{equation}
      \rho'_{(\alpha)} + (\rho_{(\alpha)} + p_{(\alpha)}) 3\cH = 0,
      \label{S:FLRW:eq:Energy-conservation}
    \end{equation}
  \item Momentum conservation
    \begin{equation}
      \partial_\eta \left[ a^4 (\rho_{(\alpha)} + p_{(\alpha)})
      \vec v_{(\alpha)}\right] = 0.
      \label{S:FLRW:eq:Momentum-conservation}
    \end{equation}
\end{itemize}
We have used $' \equiv \partial_\eta$ and $\cH\equiv a'/a$. We see that each
fluid will preserve energy and momentum separately and that $(\rho_{(\alpha)}
+p_{(\alpha)}) v_{(\alpha)}^i \propto a^{-4}$. This implies, for an equation of
state parameter which can depend on time, that
\begin{equation}
  \vec v_{(\alpha)} = \vec v_{\alpha0}\, a^{-4} \frac{1 + \omega_{\alpha0}}{1
  +\omega_\alpha(a)} \frac{\rho_{\alpha0}}{\rho_{(\alpha)}},
  \label{S:FLRW:eq:velocity-omega}
\end{equation}
where the subscript $0$ is for their present values and $\alpha = R, DR, BF,
\cdots$.  The conservation equations have the well-known solutions for a
constant equation of state parameter
\begin{align}
  \rho_{(\alpha)} =& \rho_{\alpha0} a^{-3 (1+\omega_\alpha)}
   \label{S:FLRW:eq:density-w-cte},\\
   \vec v_{(\alpha)} =& \vec v_{\alpha0} a^{3\omega_\alpha -1}.
  \label{S:FLRW:eq:velocity-w-cte}
\end{align}

Now, let us focus our attention on the effective bimetric fluid. In this case, the
energy conservation equation can be rewritten in terms of $\Omega$ using equations
(\ref{S:Gordon:eq:density}) and (\ref{S:Gordon:eq:pressure}). This leads to
\begin{equation}
   \frac{\Omega'}{\Omega} \frac{1}{1-\zeta} + \cH = 0,
   \label{S:FLRW:eq:Omega'}
\end{equation}
when $\zeta \neq 1$. As before, for the particular case of a constant function $\zeta$
this equation can be easily solved and leads to
\begin{equation}
  \Omega = \Omega_0 a^{\zeta-1}.
\end{equation}
If we split the effective fluid in its three subcomponents, the energy
conservation equation becomes
\begin{equation}
  \sum_n [\rho'_n + (\rho_n + p_n) 3\cH] = 0; ~~ n=1,2,3.
\end{equation} 
As each $\rho_n \propto \Omega^n$, each summand must vanish
individually, implying
\begin{equation}
  \rho'_n + (\rho_n + p_n) 3\cH = 0; ~~
  n=1,2,3.
  \label{S:Gordon:eq:individual-fluid-energy-conservation}
\end{equation}
That is, each component of the effective fluid conserves energy individually.
Nevertheless, as they all move with the same velocity, no such separation can
be done in the momentum conservation equation, which remains
\begin{equation}
  \partial_\eta \left[ a^4 \sum_n(\rho_n + p_n) \vec v_{BF} \right] = 0 ~~
  n=1,2,3.
  \label{S:Gordon:eq:momentum-conservation}
\end{equation}
Therefore, this effective fluid originated by the interaction between both metrics
can be interpreted as having three subcomponents, which are perfect fluids as
well, but characterized by the fact that while their energy is individually
conserved, their momenta are not. These three fluids share the same velocity
(the effective fluid velocity) which is given by equation
(\ref{S:FLRW:eq:velocity-omega}).

Finally, it must be noted that we have started by restricting attention to solutions of the form given by equations (\ref{S:FLRW:eq:g}) and (\ref{S:FLRW:eq:f}), keeping free the functions $a(t)$, $\Omega(t)$, $\zeta(t)$, and $v(t)$. We want to emphasize that we have not altered the theory considering an additional requirement, but we have focused our attention to a particular kind of solutions.
Considering the Bianchi inspired constraint, that is the conservation of the energy and momentum of the effective fluid, we are left only with two free functions, for example $\zeta(t)$ and $a(t)$. In bigravity these functions will be determined by the material content of both gravitational sectors through the equations of motion, which has to satisfy the conservation equations (\ref{S:FLRW:eq:Energy-conservation}) and (\ref{S:FLRW:eq:Momentum-conservation}), so by $\rho_m(t)$ and $\ovl\rho_m(t)$. 
On the other hand, in massive gravity the functions $\zeta(t)$ and $a(t)$ are given by the reference metric $f_{\mu\nu}$. These solutions only exist for some theories of massive gravity, those with a reference metric given by expression (\ref{S:FLRW:eq:f}) with $\Omega(t)$ and $v(t)$ fixed in terms of $\zeta(t)$ and $a(t)$ through equations equations (\ref{S:FLRW:eq:Omega'}) and (\ref{S:Gordon:eq:momentum-conservation}). Fixing $\zeta(t)$ and $a(t)$ at this point means fixing the particular massive gravity theory under investigation, being the Gordon solutions compatible only with a given matter content present in our gravitational sector that can be calculated through the equations of the dynamics. At this point we prefer to take a family of theories and not to fix the theory under investigation completely, keeping the freedom to consider different functions $\rho_m(t)$ and  $\zeta(t)$ (and, therefore, $a(t)$) of phenomenological interest.

\subsection{CMB dipole}\label{S:FLRW:SS:CMB}
Let us now study the effect of slow moving fluids over the CMB.  As shown
before, in the center of mass frame, the ordinary fluids acquire non-vanishing
velocities in order to cancel the momentum density of the bimetric fluid. This
fact implies in particular that because of the different scaling of the baryon
and radiation velocities given by equation (\ref{S:FLRW:eq:velocity-w-cte}),
after decoupling, baryon velocity will scale as $\vert \vec v_{B}\vert \propto
a^{-1}$ ($\omega_B=0$) whereas radiation velocity $\vert \vec v_R\vert=const.$
($\omega_R=1/3$). In other words, a non-vanishing $|\vec v_{BF}|$ induces a
relative motion of baryonic matter and radiation after decoupling. This motion
in turn contributes by Doppler effect to the CMB dipole anisotropy.

The photon energy measured by an observer with four velocity $U^\mu =
a^{-1}(-1,\vec{u})$ is given by \cite{Giovannini:2004rj}
\begin{equation}
  \varepsilon = U_\mu P^\mu,
\end{equation}
where 
\begin{equation}
  P^\mu = E \frac{d x^\mu}{d\lambda},
\end{equation}
is the photon four-momentum where $x^\mu(\lambda)$ is the photon geodesics,
$\lambda$ the corresponding affine parameter and $E$ parametrizes the energy.
Using the geodesics equation for the FLRW metric (\ref{S:FLRW:eq:g}), it is
easy to show that the geodesics can be written as $x^\mu = n^\mu \eta$ with
$n^\mu = (1, \vec{n})$ and $\vec{n}^2 = 1$ with $d\lambda=a^2d\eta$. Thus,  the
photon momentum results
\begin{equation}
  P^\mu = \frac{E}{a^2} n^{\mu} ,
\end{equation}
and
\begin{equation}
  \varepsilon \simeq E \left(1 + \vec{n}\vec{u} \right),
\end{equation}
to first order in velocities. The CMB temperature anisotropy is then given by 
\begin{equation}
 \frac{\delta T}{T} = \frac{a_0 \varepsilon_0 -
    a_{dec}\varepsilon_{dec}}{a_{dec}\varepsilon_{dec}} \simeq \vec{n}
    \vec{u}|^0_{dec},
\end{equation}
where the
subscripts $0$ and $dec$ refer to the present time and
decoupling, respectively. 

As mentioned before in our model baryonic matter is moving with respect to the
cosmic center of mass, as it was considered in reference \cite{Maroto:2005kc}
in a general relativistic framework, and, therefore, it will contribute to the
observer and emitter velocities. Taking into account that  $\vec{v}_B =
\vec{v}_{B0} a^{-1}$, the present value of $|\vec v_{B}|$ must be much smaller
than the one at decoupling and we can neglect it. Therefore, we have
\begin{equation}
 \frac{\delta T}{T}   \simeq \vec{n}
    \vec{v_B}|_{dec},
\end{equation}
which is just a dipole contribution. The experimental amplitude for the CMB
dipole is \cite{Fixsen:1993rd}
\begin{equation}
  \frac{\delta T}{T}\bigg |_{dipole} = 1.23 \times 10^{-3},
\end{equation}
so that we can set a limit on the baryonic velocity at decouplig 
$\vert \vec v_{B,dec}| < 1.23 \times 10^{-3}$, which implies that today $\vert
\vec v_{B0}| < 1.1 \times 10^{-6}$. 
In addition, since prior to decoupling baryonic matter and radiation should
have shared the same velocity, we have $\vec v_R \simeq \vec v_{B,dec}$ and,
therefore, $|\vec v_R| < 1.23\times 10^{-3}$. 

\subsection{Monocomponent bimetric fluid}
\label{S:FLRW:SS:Monocomponent}
In this section we will study the case when the effective fluid is
composed only of one component, i.e.  only one non-vanishing 
$\beta_n$. Then, this fluid will be identified with one
of the dark fluids existing in our universe. In fact, as there is only a
component, it is equivalent to work with $\zeta$ and $\Omega$ or $\omega$ and
$\rho$, since they are related by
\begin{align}
  1-\zeta = 3(\omega_n + 1)/n,
  \label{S:FLRW:SS:Monocomponent:eq:zeta-w}\\
  \rho_n = \rho_{n0} \left( \Omega/\Omega_0 \right)^{n},
  \label{S:FLRW:SS:Monocomponent:eq:rho-W}
\end{align}
where the subscript $0$ stands for the present value of the function, and
where  the other constants present in equations (\ref{S:Gordon:eq:fluid3})
have been absorbed into $\rho_{n0}$.
For the sake of simplicity, we will assume that all the fluids move along the 
x-axis so that $\vec v_{(\alpha)}=(v_{(\alpha)},0,0)$.
We can express the spatial velocity of the effective fluid,  $v_{BF}=v_n$, in terms of
$\zeta$ and $\Omega$, using the fact that $(\rho_n + p_n)v_n \propto
a^{-4}$, as shown in equation (\ref{S:FLRW:eq:Momentum-conservation}). This is
\begin{equation}
  v_n = v_{n0}\, a^{-4} \frac{1 - \zeta_0}{1 - \zeta(a)}
  \left(\frac{\Omega_0}{\Omega}\right)^n.
\end{equation}
It must be emphasized that the particular form of the functions $\zeta(\eta)$
and $\Omega(\eta)$ will be fixed by the theory itself in case of massive
gravity since $f_{\mu\nu}$ in equation (\ref{S:FLRW:eq:f}) is given a priori.
In contrast, for bigravity, these functions can be obtained from the modified
Einstein equations of the $f$-sector, equation (\ref{S:notes:eq:Einstein-f}),
once the material content coupled to that sector is fixed. 

In the following, we will consider some phenomenological interesting cases
assuming that the reference metric or the material content of the $f$-sector is
such that it gives compatible parameters with our calculations in massive
gravity and bigravity, respectively.  Hence, considering that the function
$\zeta$ is constant, which implies $\omega_n = \cst$, we have
\begin{align}
  \rho_n = \rho_{n0} a^{-3(1+\omega_n)} &~\mbox{ or }~ \Omega = \Omega_0
a^{\zeta-1},
\label{S:FLRW:SS:Monocomponent:eq:density}\\
v_n = v_{n0} a^{3\omega_n -1} ~&\mbox{ or }~  v_n = v_{n0} a^{n(1-\zeta)-4}.
\label{S:FLRW:SS:Monocomponent:eq:velocity}
\end{align}
Note that a cosmological constant fluid can already be obtained in bigravity
or massive gravity from the contributions of $\beta_0$ and $\beta_4$ to $g$-sector
and $f$-sector, respectively. Therefore, we will not consider  the  case 
($\zeta=1$) in which the bimetric fluid behaves as cosmological constant.

\subsubsection{Bimetric dark radiation}\label{S:FLRW:SS:Monocomponent:SSS:Radiation}
Let us consider that the monocomponent bimetric fluid corresponds to dark
radiation, while the other cosmic fluids (radiation, dark and baryonic matter
and cosmological constant) will be described by $T\mmn$. Taking into account
equation (\ref{S:FLRW:SS:Monocomponent:eq:zeta-w}), it can be noted that in
order to describe a bimetric fluid with $\omega_n =1/3$, $\zeta$ has to take
negative values: $\zeta = -3$ if we want to have dark radiation originated by
component $n=1$, $\zeta = -1$ for $n=2$, and $\zeta=-1/3$ for $n=3$.  Moreover,
from equation (\ref{S:FLRW:SS:Monocomponent:eq:density}) one has $\Omega =
\Omega_0 a^{-4/n}$.  We will work with $\rho$ and $\omega$ instead of $\Omega$
and $\zeta$ for the following calculations, since they are the physical
observables.

Using the gauge choice $\vec S = \vec 0$, it is possible to find a relation
between the different fluids velocities. From equation (\ref{S:FLRW:eq:S}) this
is
\begin{equation}
  \frac{4}{3} \Omega_{DR} v_{DR} + \Omega_{B} v_{B0} + \Omega_{DM}
  v_{DM0} + \frac{4}{3} \Omega_R  v_R = 0,
  \label{S:FLRW:SS:Monocomponent:SSS:Radiation:eq:momentum-conservation}
\end{equation}
where the $\Omega_{x}$ are density parameters and $DR$, $R$, $B$ and
$DM$ stand for dark radiation, radiation, baryonic matter and dark matter,
respectively. It is important to remark that an exact cosmological constant
fluid does not contribute to the momentum density and, therefore, it does not enter
in momentum conservation. 

\begin{figure}[htb]
  \centering
  \includegraphics[width=.6\textwidth]{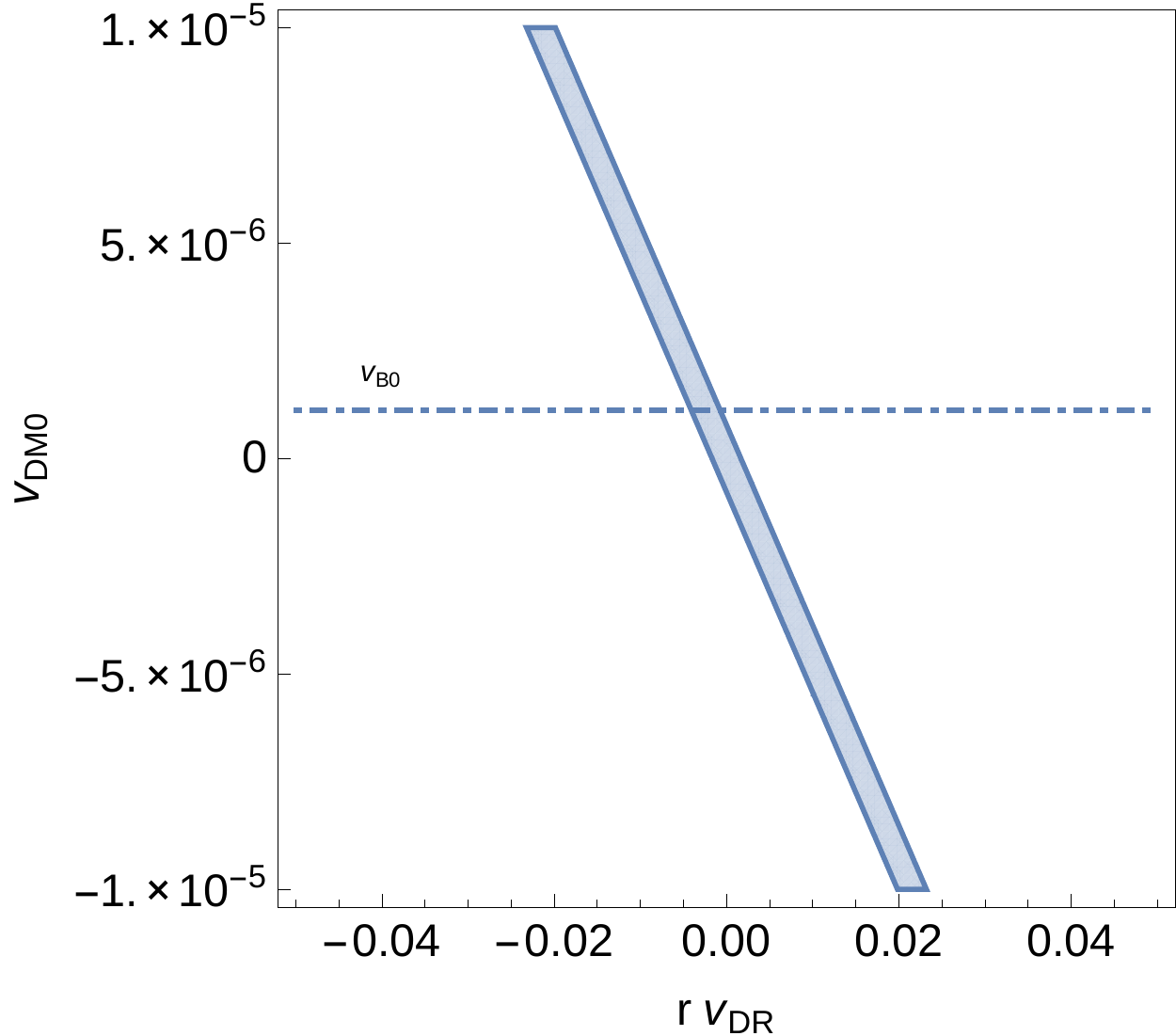}
  \caption{Compatible values of dark matter and dark radiation velocity
  with the CMB dipole. The horizontal line is the present velocity of baryonic
  matter. Remember $r<0.1$, as $\Omega_{DR} = r \,\Omega_R$}
\label{S:FLRW:SS:Monocomponent:fig:radiation-general}
\end{figure}

Imposing the additional conditions  $\vert v_R\vert < 1.23 \times 10^{-3}$
and $\vert v_{B0}| < 1.1 \times 10^{-6}$,  we obtain a
relation between dark matter and dark radiation velocities. In figure
\ref{S:FLRW:SS:Monocomponent:fig:radiation-general}, the allowed values of
$v_{DR}$ and $v_{DM0}$ have been plotted. Dark matter velocity is seen to be
able to vary only in a range of order $10^{-8}$ for a given value of $r
v_{DR}$, with $r$ that fraction of dark radiation with respect to total
radiation density ($\Omega_{DR} = r\, \Omega_R$, $r< 0.1$ \cite{Ade:2015xua}).

Within the range of  solutions compatible with CMB dipole observations shown in
figure \ref{S:FLRW:SS:Monocomponent:fig:radiation-general}, we will concentrate
in the case in which dark matter is at rest with respect to the cosmic center
of mass. In the standard thermal dark matter
scenario, dark matter decoupled from radiation and baryonic matter in the early
universe, and its velocity would have been falling as $a^{-1}$ since then,
making it negligible at present.
Then, if we take $v_{DM0} \simeq 0$ together with $|v_R| < 1.23 \times 10^{-3}$
and $|v_{B0}| < 1.1 \times 10^{-6}$, the momentum conservation equation
(\ref{S:FLRW:SS:Monocomponent:SSS:Radiation:eq:momentum-conservation}) yields
\begin{equation}
  v_{DR} = -r^{-1}\left[\frac{3}{4} \frac{\Omega_{B}}{\Omega_R} v_{B0} + v_R \right], 
\end{equation}
and 
\begin{equation}
  |v_{DR}| < 1.8 \times 10^{-2}.
\end{equation}
Thus, we see that in this scenario, a dark radiation bimetric fluid with
a non-negligible velocity is still compatible with observations.

\begin{figure}[htb]
  \centering
  \begin{subfigure}[b]{0.49\textwidth}
    \includegraphics[width=\textwidth]{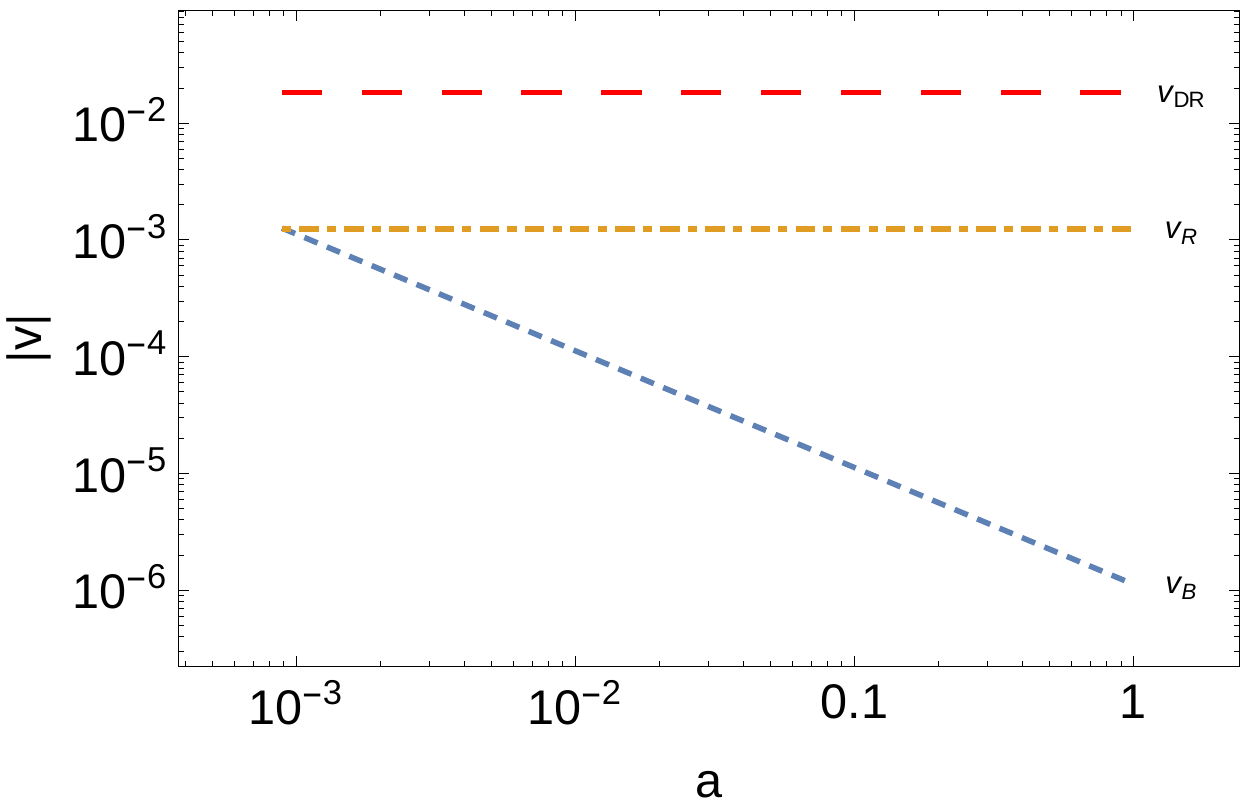}
  \end{subfigure}
  \begin{subfigure}[b]{0.49\textwidth}
    \includegraphics[width=\textwidth]{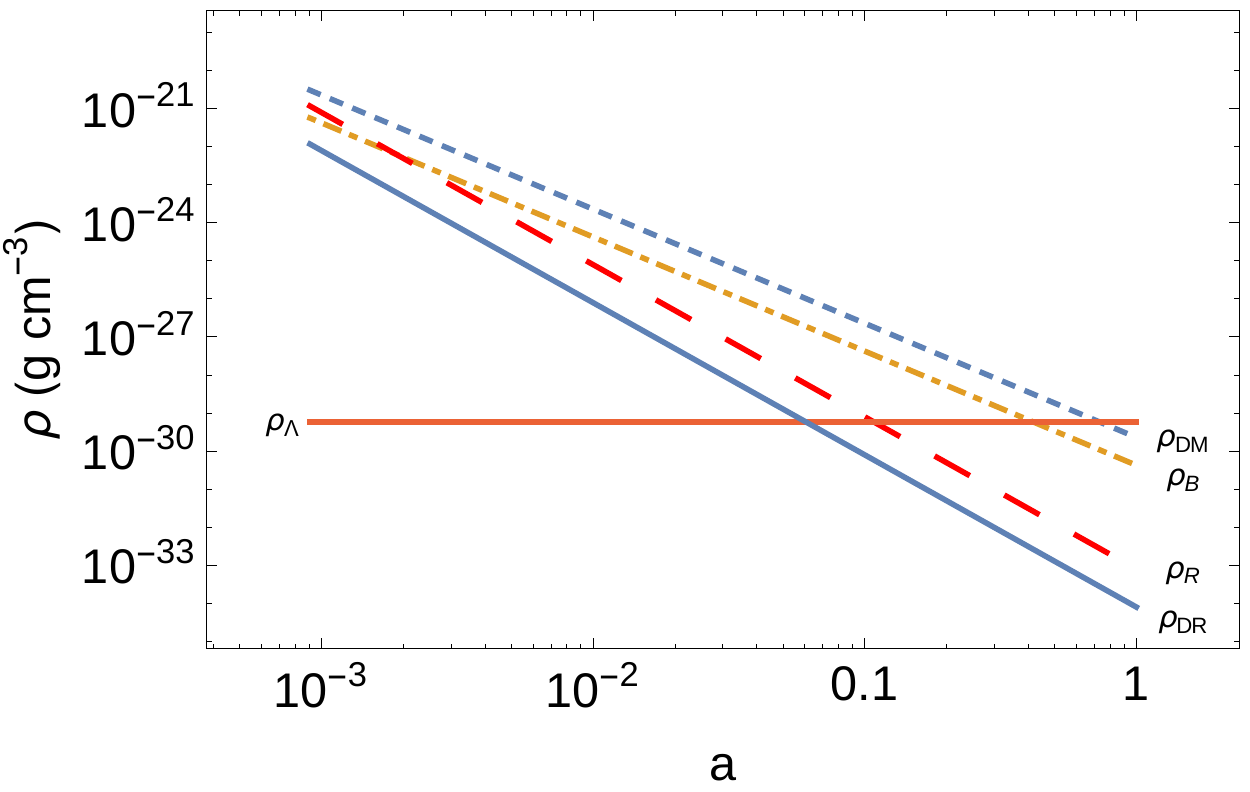}
  \end{subfigure}
  \caption{Velocities (left) and densities (right) for a universe filled with
  dark and usual radiation, dark and baryonic matter and a cosmological
  constant. Their present values have been taken to be compatible with the CMB
  dipole. Note that $v_{DR}$ has been obtained supposing dark matter is at
  rest with respect to the cosmic center of mass. In addition, it has
  been taken that $\Omega_R^{dark}= 0.1 \,\Omega_R$. It should be noted that for
  early times the linear approximation breaks and our treatment is not
valid.}
\label{S:FLRW:SS:Monocomponent:fig:radiation-3}
\end{figure}

In figure \ref{S:FLRW:SS:Monocomponent:fig:radiation-3}, the evolution of the
fluids velocities (except for cosmological constant, that is irrelevant) and
densities has been plotted for the case that saturates all inequalities.  It
should be noted that, as we are working linearly in velocities, they will not
be valid for arbitrary early times; whereas in the future our approximation
will become even better. 

\subsubsection{Bimetric dark matter}
\label{S:FLRW:SS:Monocomponent:SSS:Matter}
Let us now identify the bimetric fluid with dark matter. As $\omega_n = 0$,
from equation (\ref{S:FLRW:SS:Monocomponent:eq:zeta-w}) we get $\zeta = -2$ if
only  $n=1$ fluid is present, $\zeta = -1/2$ for  $n=2$, and $\zeta=0$ for
$n=3$. The only possible purely matter fluid would be due to component 1 or 2,
since third component is forbidden by the condition $\zeta\neq 0$ to preserve
both metrics Lorentzian. We can use, nevertheless, the positive square root
branch of $\zeta$ with fluid 3 if we  take $\zeta$ arbitrarily near to zero,
instead of using the exact matter solution.  From
(\ref{S:FLRW:SS:Monocomponent:eq:density}) we get in this case $\Omega =
\Omega_0 a^{-3/n}$.

 We assume that the universe is filled with radiation, dark and baryonic matter
 and cosmological constant. From the previous section we know that $v_R$
and $v_{B0}$ are already fixed by dipole with values $|v_R| < 1.23\times
10^{-3}$ and $|v_{B0}| < 1.1 \times 10^{-6}$. Therefore, we only need to
calculate the velocity of dark matter. Using momentum conservation and $\vec S
= \vec 0$, we have
\begin{equation}
  \frac{4}{3}\Omega_R v_R + \Omega_{DM} v_{DM0} + \Omega_B v_{B0} = 0,
  \label{S:FLRW:SS:Monocomponent:eq:matter}
\end{equation}
yielding, 
\begin{equation}
  v_{DM0} = -\frac{4}{3} \frac{\Omega_R}{\Omega_{DM}}v_R -
  \frac{\Omega_B}{\Omega_{DM}} v_{B0},
\end{equation}
and
\begin{equation}
  |v_{DM0}| < 6.9 \times 10^{-7}.
\end{equation}

\subsubsection{Interpolating bimetric dark fluid}
\begin{figure}[htb]
  \centering
  \includegraphics[width=0.6\textwidth]{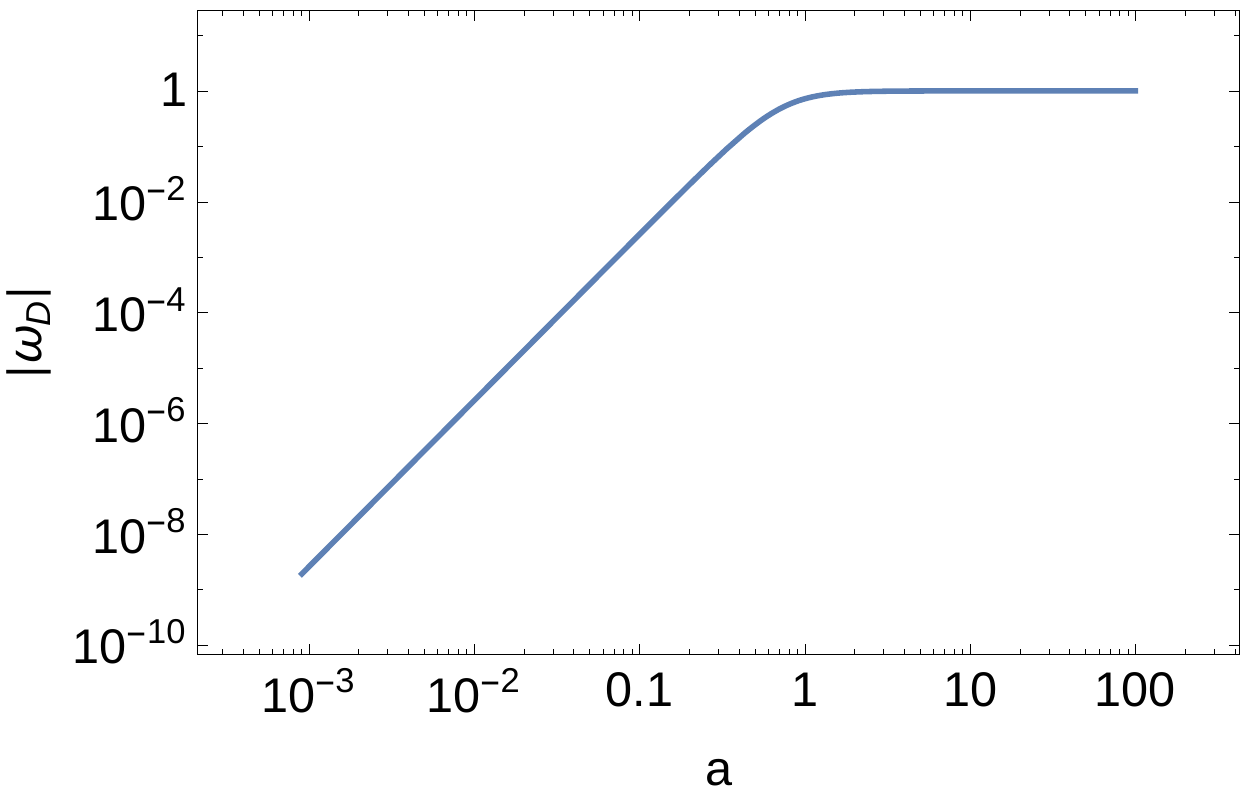}
  \caption{Equation of state parameter for a fluid that evolves in time from dark
  matter to cosmological constant.}
  \label{S:FLRW:SS:Monocomponent:fig:interpolation-w}
\end{figure}

We consider now the case in which dark matter and dark energy are indeed the
same fluid, which we identify with the bimetric fluid.  We parametrize the
equation of state parameter of this interpolating bimetric dark fluid as
\begin{equation}
  \omega_{D}(a) = \frac{p_{DM}+p_\Lambda}{\rho_{DM} + \rho_\Lambda} = 
  -\frac{1}{1+ \Omega_{DM} a^{-3}/\Omega_\Lambda},
  \label{S:FLRW:SS:Monocomponent:eq:interpolation-w}
\end{equation}
where the subscript $D$ stands for ``dark''. This equation of state parameter
has been plotted in figure \ref{S:FLRW:SS:Monocomponent:fig:interpolation-w}.
It must be noted that the only bimetric fluid component able to evolve
in such a way is the third one, since the others cannot avoid the singular
behavior at $\zeta=0$, then $\beta_3\neq0$. Therefore, from equation
(\ref{S:Gordon:eq:fluid3}), $\zeta = -\omega_D$ and, from equation
(\ref{S:FLRW:eq:Omega'}), $\Omega = \Omega_0 \left(\frac{\Omega_\Lambda +
  \Omega_{DM} a^{-3}}{\Omega_\Lambda + \Omega_{DM}}\right)^{1/n}$.

\begin{figure}[htb]
  \centering
  \begin{subfigure}[b]{0.49\textwidth}
    \includegraphics[width=\textwidth]{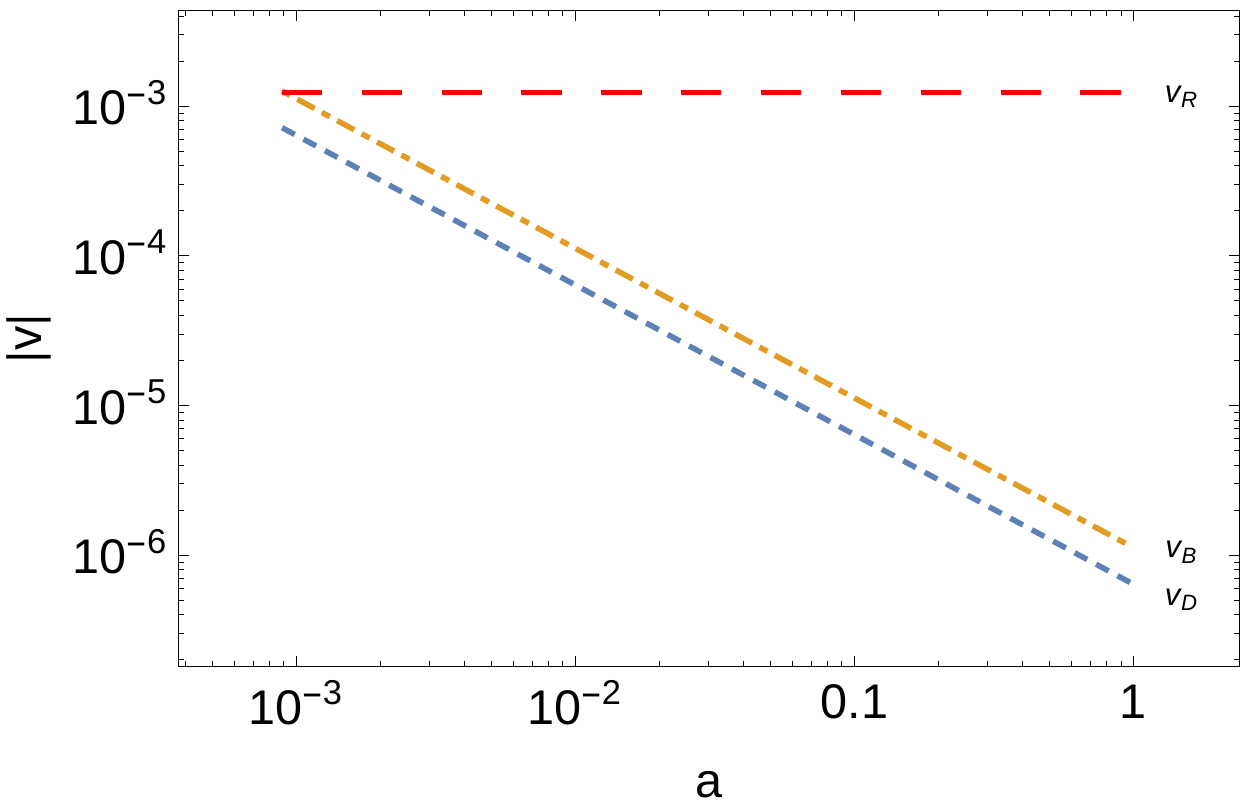}
  \end{subfigure}
  \begin{subfigure}[b]{0.49\textwidth}
    \includegraphics[width=\textwidth]{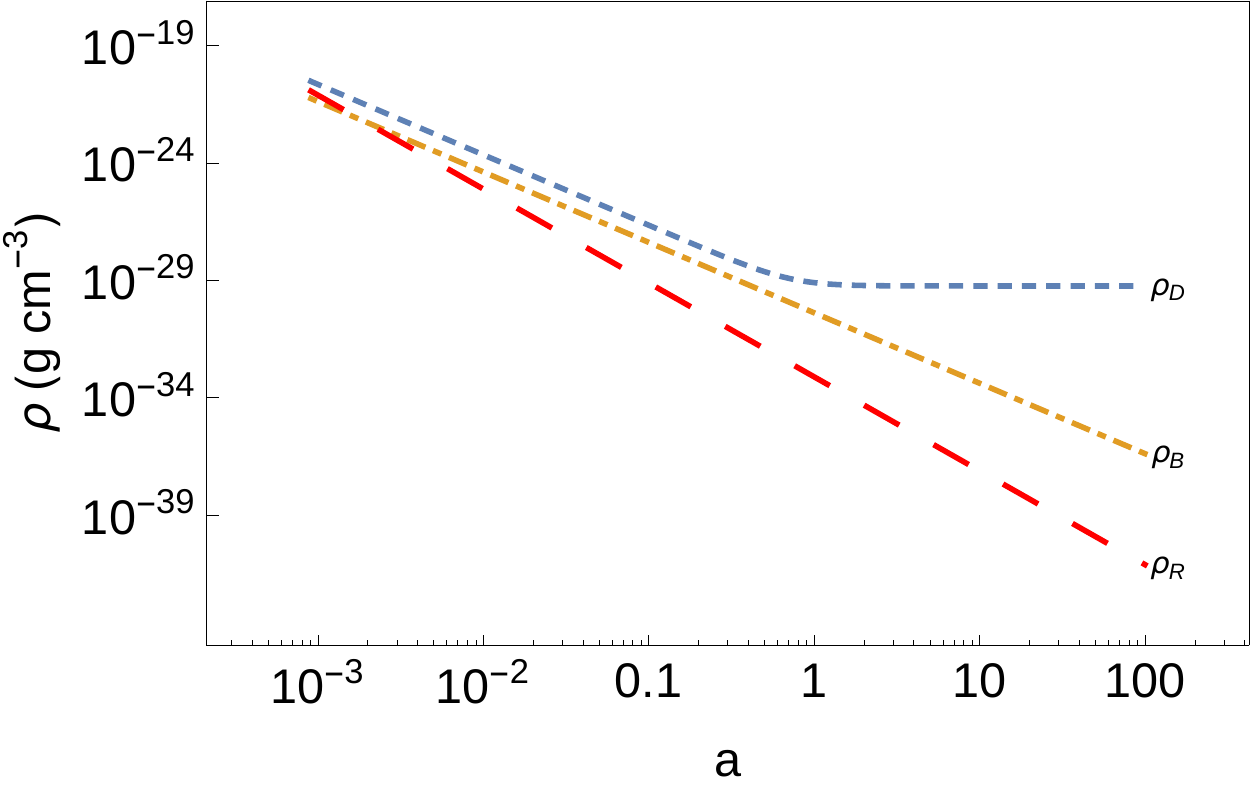}
  \end{subfigure}
  \caption{Velocities (left) and densities (right) in a universe with
    radiation, baryonic matter and a fluid that evolves in time from dark
    matter to cosmological constant (subscript $D$). The present values have
  been taken to be compatible with the CMB dipole.}
  \label{S:FLRW:SS:Monocomponent:fig:interpolation}
\end{figure}

In the first place, we solve the conservation equations for such a fluid,
equations (\ref{S:FLRW:eq:Energy-conservation}) and
(\ref{S:FLRW:eq:Momentum-conservation}), to obtain
\begin{align}
  \rho_D =& \rho_\Lambda + \rho_{DM0} a^{-3},\\
  v_D =& v_{D0} a^{-1},
\end{align}
It should be noted that the velocity of this interpolating fluid will always
behave as that of matter even though at late times the fluid behaves as a
cosmological constant.  This is so because the cosmological constant does not
contribute to the momentum density as  mentioned before. In the second place,
we consider again the condition $\vec S = \vec 0$ to fix the present values of
velocities. This reads 
\begin{equation}
  (1+\omega_D)\Omega_D v_{D0} + \frac{4}{3} \Omega_R v_R + \Omega_B v_{B0} = 0.
\end{equation}
So, the present velocity for the dark sector is
\begin{equation}
  v_{D0} = - \frac{\frac{4}{3}\Omega_R v_R + \Omega_B v_{B0}}{(1+\omega_D)\Omega_D},
\end{equation}
yielding
\begin{equation}
  |v_{D0}| < 6.4 \times 10^{-7}.
\end{equation}
The energy densities and the absolute values of the velocities have been
plotted in figure \ref{S:FLRW:SS:Monocomponent:fig:interpolation}, considering
their maximum possible values.

Finally, it should be noted that, since $\rho_{n0} \propto \beta_n \Omega_0$ we
cannot fix, for a monocomponent fluid at least, both $\Omega_0$ and $\beta_n$
from observations. 

\subsection{Multicomponent bimetric fluid}\label{S:FLRW:SS:Multicomponent}
As we have seen in section \ref{S:Gordon:SS:split}, the effective fluid can be
split in three subcomponents. In the previous section we have studied the case
when only one of those subcomponents was present. Now, we will describe
qualitatively two cases when there is more than one component.

It is easy to see from the definition of $\rho_{BF}$ and $\rho_n$, equations
(\ref{S:Gordon:eq:density}) and (\ref{S:Gordon:eq:fluid3}), that when $\Omega$
depends on time and we have several subcomponents, the bimetric fluid energy
density will exhibit the behavior of the dominant subcomponent at every time.
This change in the evolution will be reflected also in  the velocity through
equation (\ref{S:FLRW:eq:velocity-omega}).  Before studying some interesting
cases, we want to emphasize again that when we refer to fixing the theory
functions $\Omega$ and $\zeta$, we actually mean that we choose a theory of
massive gravity with a reference metric $f_{\mu\nu}$ given by equation
(\ref{S:FLRW:eq:f}) with those functions, or that we consider that the material
content of the $f$-sector is such that it leads to a $f_{\mu\nu}$ metric
corresponding to those functions in bigravity.

\begin{figure}[htb]
  \centering
  \begin{subfigure}[b]{0.49\textwidth}
    \includegraphics[width=\textwidth]{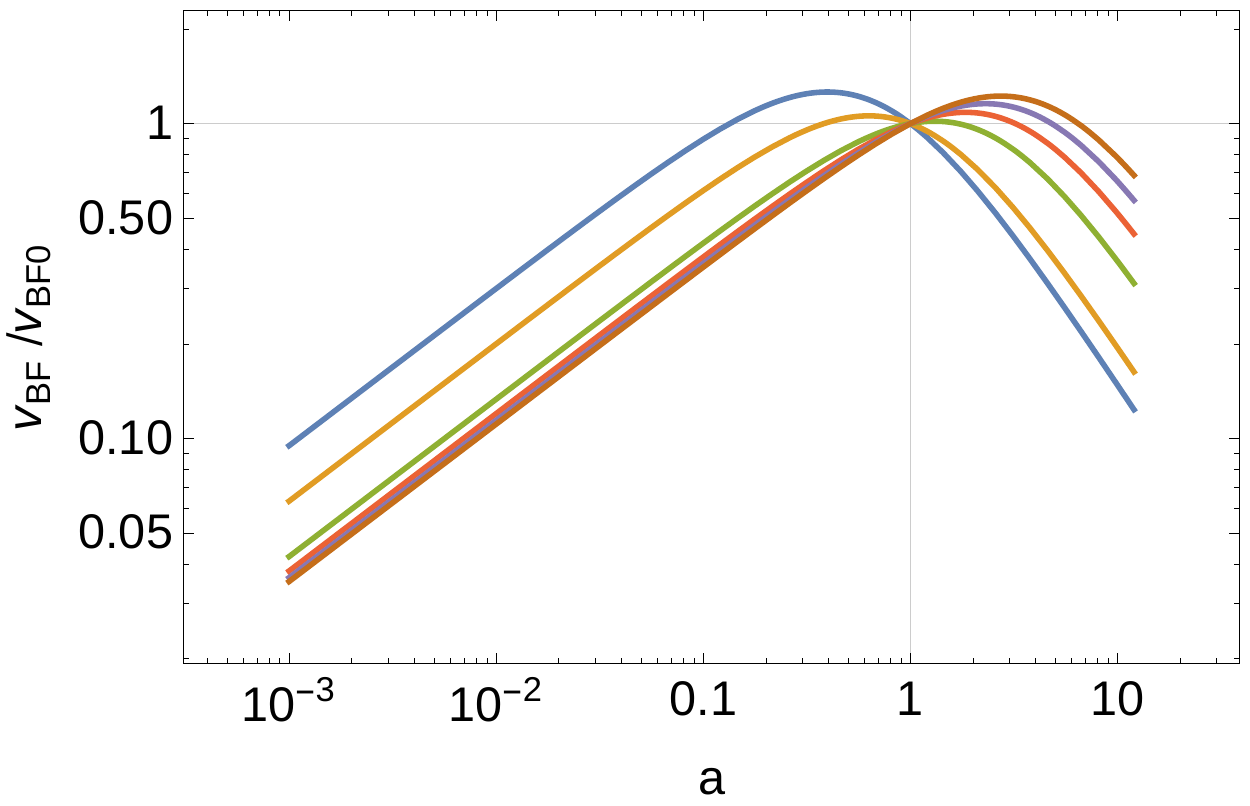}
    \end{subfigure}
  \begin{subfigure}[b]{0.49\textwidth}
    \includegraphics[width=\textwidth]{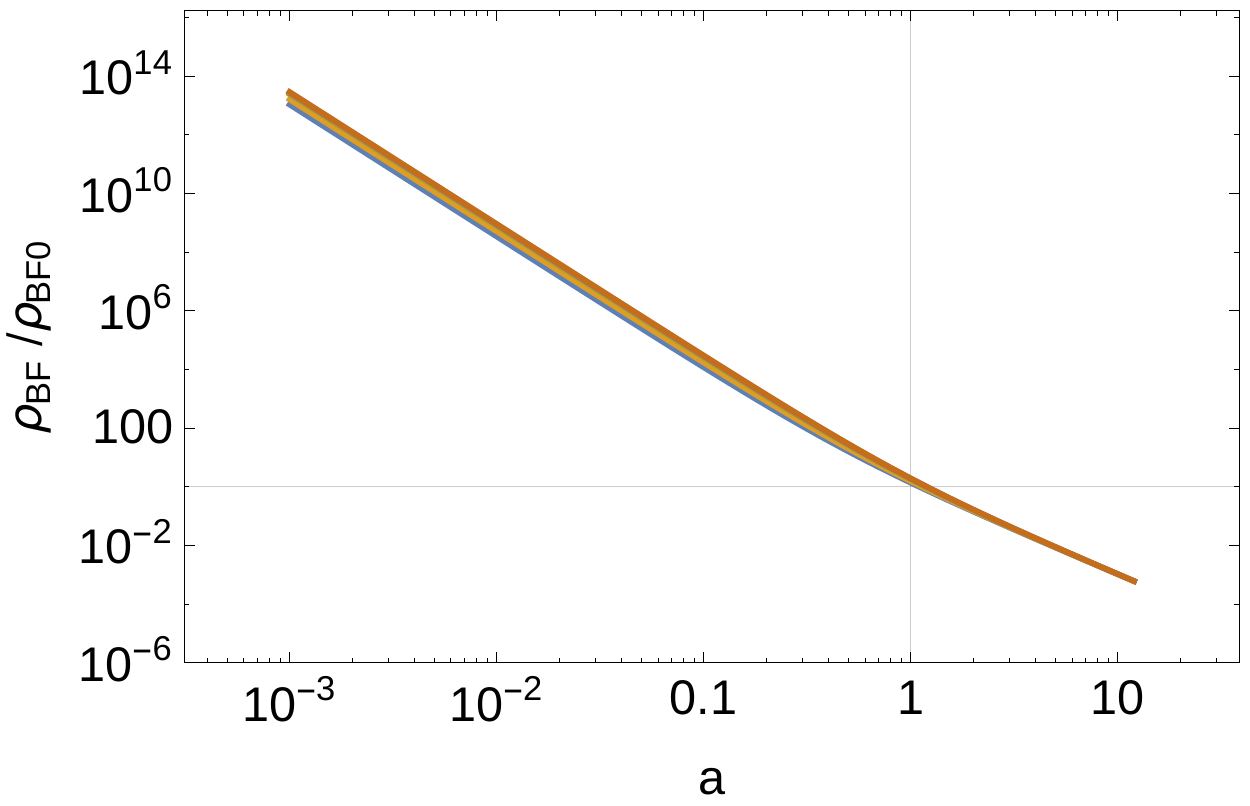}
    \end{subfigure}
    \caption{Effective fluid velocity (left) and density (right) functional
      behavior in $a$ and $r$ for $\zeta = -1/2$. Here, we have plotted
      $r=0.5,1,3,5,7,9$ from top to bottom. We observe for velocity that the
      lower $r$, the higher the peak and the acceleration.  There is also
      a shift to the left as the value of $r$ decreases. All the energy densities have almost the same behavior.}
      \label{S:FLRW:SS:Multicomponent:fig:z=-1/2}
\end{figure}

Now, let us consider an example in which we have a matter behaviour at late time
with an early phase in which the bimetric fluid is self-accelerated. 
For that purpose we need a bimetric fluid with only components 2 and 3 and with
$\zeta = -1/2$. Then, from the equation
(\ref{S:FLRW:SS:Monocomponent:eq:zeta-w}), we have $\omega_2=0$ and $\omega_3 =
1/2$. Taking into account equation (\ref{S:Gordon:eq:momentum-conservation}),
the velocity is given by
\begin{equation}
   v_{BF} = v_{BF0} (1+r) (a + r/\sqrt{a})^{-1},
\end{equation}
with $r=\Omega_0 \beta_3/\beta_2$. Note that we cannot constrain simultaneously
both constants, $v_{BF0}$ and $r$ with the CMB dipole. For this reason we
have plotted only its functional behavior for different values of $r$ in figure
\ref{S:FLRW:SS:Multicomponent:fig:z=-1/2}. It can be seen that the fluid
accelerates at short times. The maximum velocity is found at
$a = (r/2)^{2/3}$ with $v \propto r^{-2/3}$; i.e. the lower the $r$ the higher
the peak.  The density associated to this effective fluid is
\begin{equation}
  \frac{\rho_{BF}}{\rho_{BF0}} = a^{-3} +  \frac{r}{1+r} a^{-9/2},
\end{equation}
the matter contribution dominates at late times whereas at early time a 
decrease faster than radiation is exhibited. The dependence on $r$ is highly
suppressed on time (see figure \ref{S:FLRW:SS:Multicomponent:fig:z=-1/2}).

\begin{figure}[htb]
  \centering
  \begin{subfigure}[b]{0.49\textwidth}
    \includegraphics[width=\textwidth]{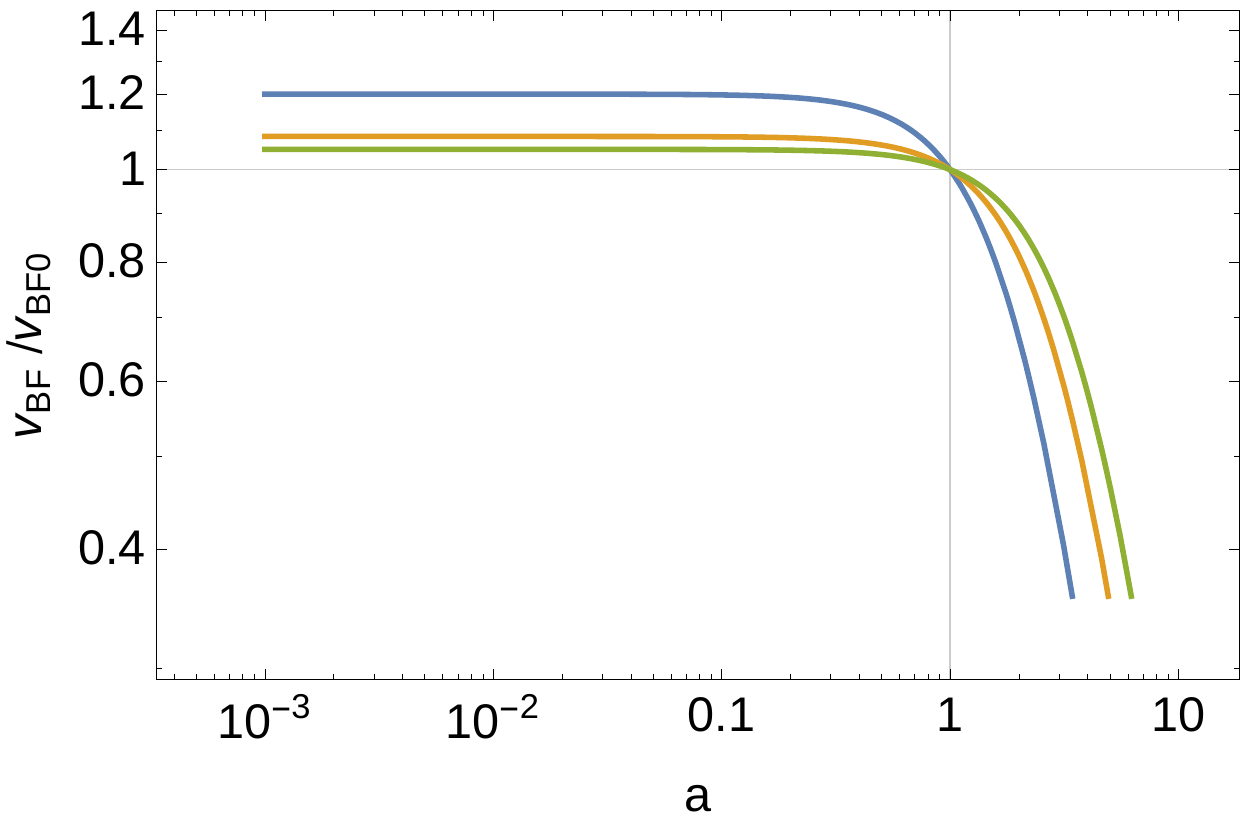}
  \end{subfigure}
  \begin{subfigure}[b]{.49\textwidth}
    \includegraphics[width=\textwidth]{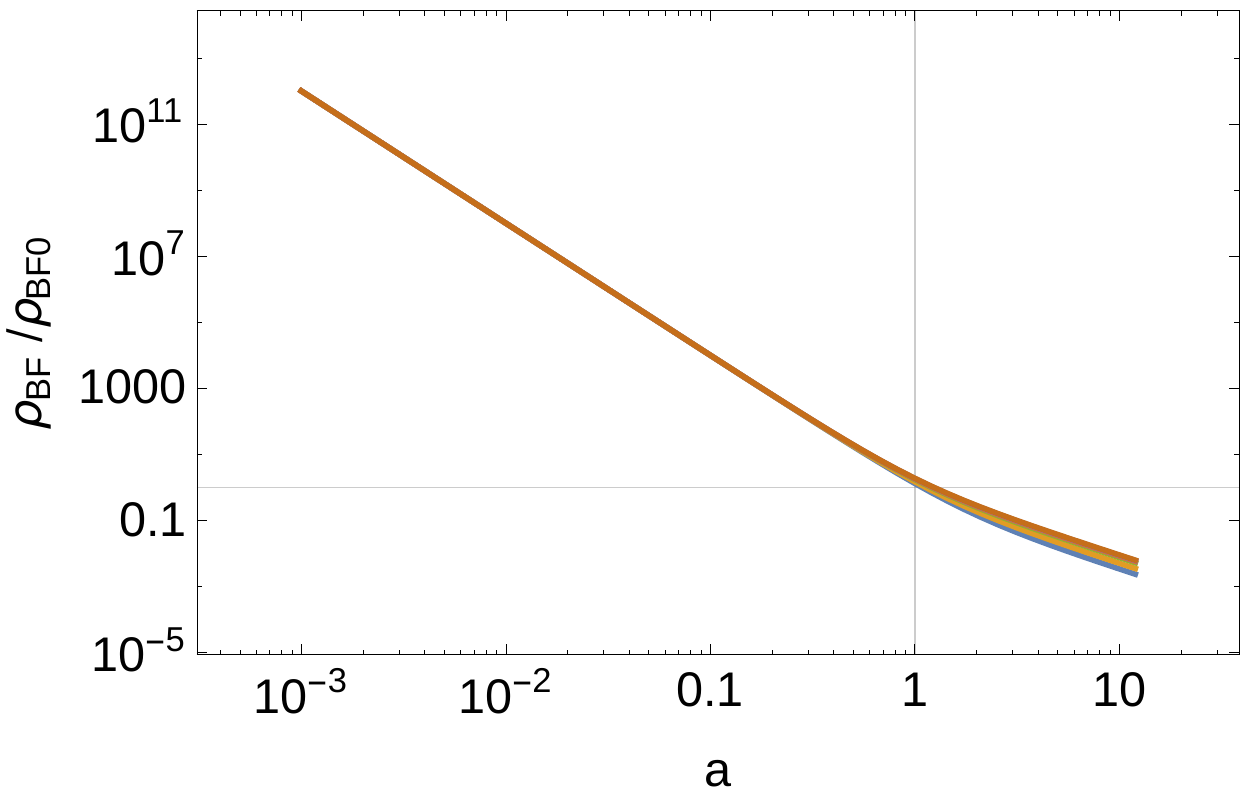}
  \end{subfigure}
    \caption{Effective fluid velocity (left) and density (right) functional
      behavior when the second component is radiation and the third is turned
      off. It is clearly monotonically decreasing and all converge to 0 for
      large $a$ and $r$. We have plotted the cases $r=5,12,20$, listed from top to bottom.}
      \label{S:FLRW:SS:Multicomponent:fig:z=-1}
\end{figure}

Other interesting situation is that when $\zeta = -1$ and, therefore, $\omega_1
= -1/3$ and $\omega_2 = 1/3$; therefore, the second component of the effective
fluid behaves as radiation. Now, the velocity of the effective fluid is given
by
\begin{equation}
  v_{BF} = v_{BF0} (1+r) (a^2 + r)^{-1},
\end{equation}
and its density
\begin{equation}
  \frac{\rho_{BF}}{\rho_{BF0}} = a^{-4} + \frac{r}{1+r} a^{-2},
\end{equation}
with $r = \Omega_0 \beta_2/\beta_1$. These functions have been plotted in
figure \ref{S:FLRW:SS:Multicomponent:fig:z=-1}. This time we see that the
velocity is constant up to some time when it starts to decrease. Its energy
density is again almost unaffected by the value of $r$. As in the previous
case, the bimetric fluid can be split into two contributions, one that behaves
as radiation, which dominates at early times, and another driving the
cosmological evolution at late times.

More general situations with all three subcomponents could be explored.
Nevertheless, the qualitative behaviour  will be similar to those shown in this
section.

\section{Null bimetric fluid}\label{S:Kerr}
In this section, we will restrict our attention to solutions of massive gravity
and bigravity with metrics that are related through a different kind of causal
relation. We consider the Kerr-Schild ansatz which assumes that both metrics
have one null vector in common. This ansatz can be expressed as
\cite{Baccetti:2012ge}
\begin{equation}\label{KS}
  f_{\mu\nu} = \Omega^2 (g_{\mu\nu} + \xi\, l_\mu l_\nu)
\end{equation}
where $l_\mu$ is the common null vector and $\xi$ is a function that can take
any value. It must be noted that for $\xi = 0$ we recover a conformal
transformation, whose effect for the bimetric theory is that of a cosmological
constant.  When considering the Kerr-Schild ansatz (\ref{KS}) it can be seen
that \cite{Baccetti:2012ge}
\begin{equation}
  \gamma\mn= \Omega \left( \delta\mn + \xi\, l^\mu l_\nu \right),
\end{equation}
The stress-energy tensor of the effective fluid obtained in reference
\cite{Baccetti:2012ge} can be expressed as
\begin{equation}
  T\mn = (\rho_N + p_N) l^\mu l_\nu + p_N \delta\mn,
  \label{S:Kerr:eq:T-null}
\end{equation}
where $N$ stands for bimetric null fluid and the density and pressure are given
by
\begin{align}\label{S:Kerr:eq:rho}
 \rho_N =& \frac{\Omega}{2}\left[ \beta_1(6 -\xi) + 2 \beta_2 \Omega (3 -\xi) + \beta_3 \Omega^2 (2 - \xi)\right],\\
  p_N =& -\Omega(3\beta_1 + 3\Omega \beta_2 + \beta_3 \Omega^2),
  \label{S:Kerr:eq:preassure}
\end{align}
and whose sum takes the form
\begin{equation}
  \rho_N + p_N = - \frac{\Omega}{2} \xi (\beta_1 + 2 \Omega \beta_2 + \Omega^2
  \beta_3),
  \label{S:Kerr:eq:density+preassure}
\end{equation}
which will be useful in the next section to obtain $\xi$.  On the other hand,
for the other sector one has
\begin{equation}
  \ovl T^\mu _\nu = (\ovl \rho_N + \ovl p_N) l^\mu l_\nu + \ovl p_N \delta^\mu_\nu
\end{equation}
where
\begin{align}
  \ovl \rho_N =& \frac{1}{2\Omega^3} \left[ \beta_1 (2+\xi) +2 \beta_2 \Omega
  (3+\xi) + \beta_3 \Omega^2 (2+\xi)\right],\\
  \ovl p_N =& -\frac{1}{\Omega^3} (\beta_1 + 3\beta_2 \Omega + \beta_3
  \Omega^2).
\end{align}
Thus, we see that unlike the generalized Gordon ansatz which was related to perfect fluids with 
time-like velocities, the Kerr-Schild ansatz introduces an effective perfect
fluid with null velocity.  It should be noted that whereas in reference
\cite{Baccetti:2012ge} $\rho_N+p_N$ has been interpreted as the flux of the
null fluid, here we have chosen this formulation to emphasize that, in contrast
to the previous section, we are now considering the high velocity limit.

\subsection{Interacting effective null fluids}\label{S:Kerr:SS:split}
\begin{figure}[htb]
  \begin{subfigure}[b]{0.49\textwidth}
  \centering
  \includegraphics[width=\textwidth]{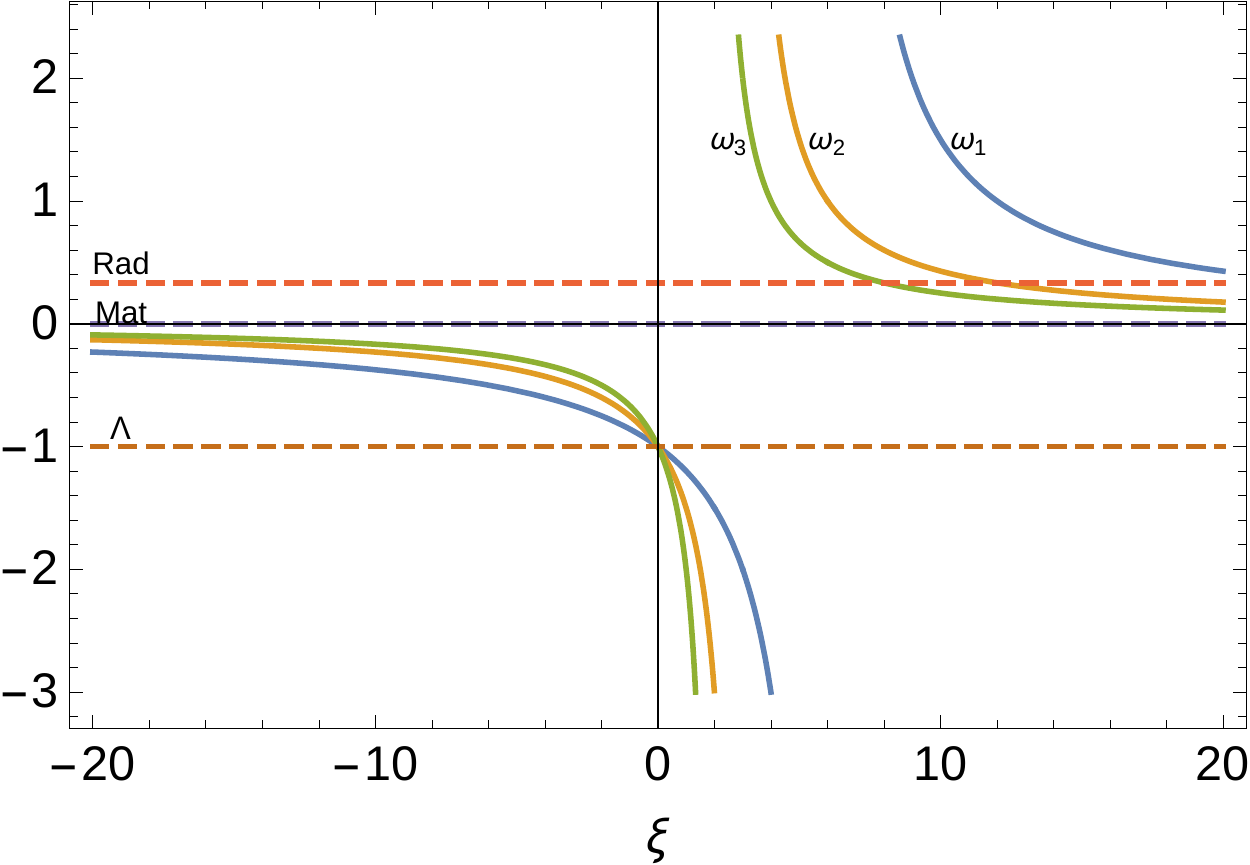}
  \end{subfigure}
  \begin{subfigure}[b]{0.49\textwidth}
    \includegraphics[width=\textwidth]{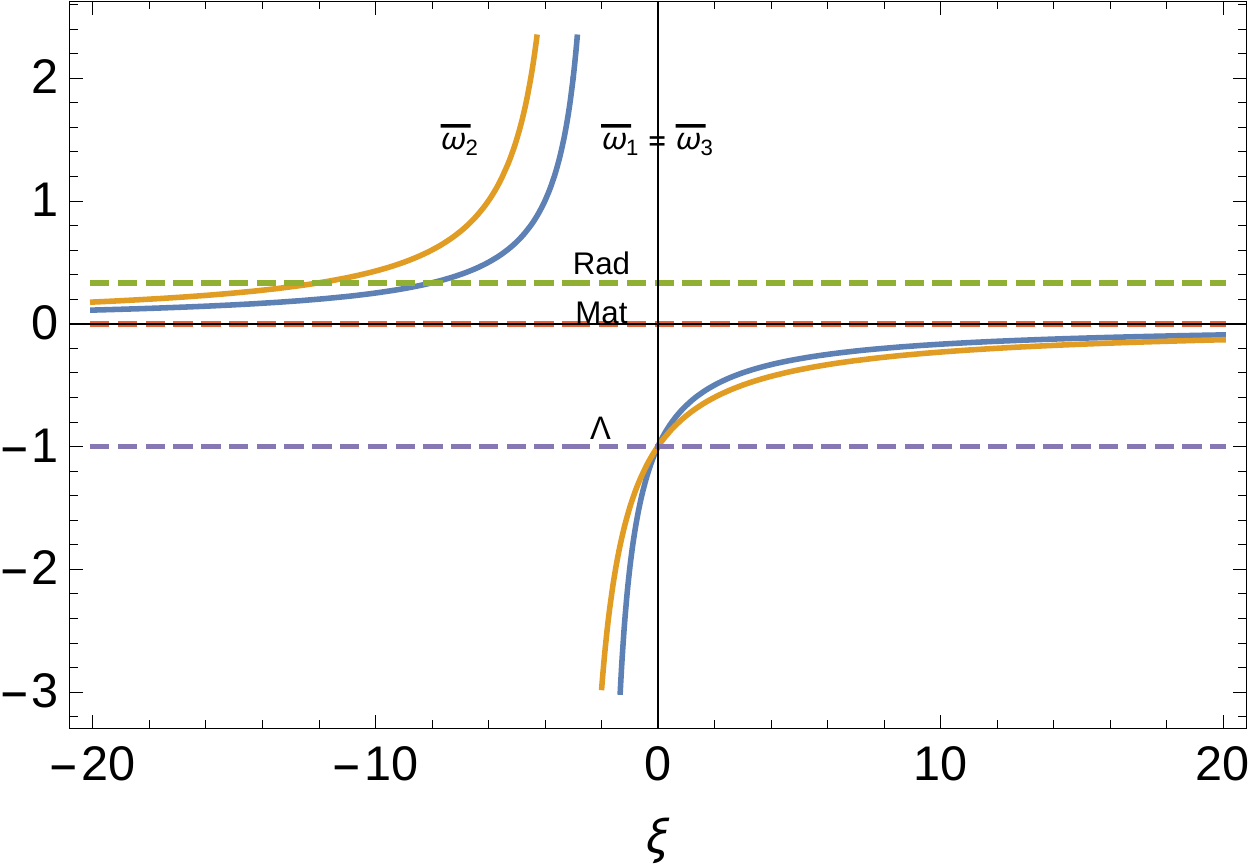}
  \end{subfigure}
  \caption{Equation of state parameter, seen from the $g$-sector (left) and the
    $f$-sector (right), for each fluid as a function of $\zeta$. The vertical
    lines show where the divergences take place. These divergences cause no
    problem since they are canceled when multiplied by $\rho$ to obtain
    $p$. Equivalently, they are harmless since they do not appear in
  $\rho$ nor $p$.} 
    \label{S:Kerr:fig:plotW-z}
\end{figure}

As in the Gordon  case, we can split the bimetric fluid in three subcomponents.
For our gravitational sector we have
\begin{align}\nonumber
  \rho_1 &=  \frac{1}{2}\beta_1 \Omega (6 -\xi), &p_1 &=  -3\beta_1
  \Omega, &\omega_1 &=  -\frac{6}{6-\xi};\\\nonumber
  \rho_2 &=  \beta_2 \Omega ^2 (3-\xi), &p_2 &= - 3\beta_2 \Omega^2,
   &\omega_2 &=  -\frac{3}{3 - \xi};\\
  \rho_3 &=  \frac{1}{2}\beta_3 \Omega^3 (2-\xi), &p_3 &=  -\beta_3 \Omega^3,
  &\omega_3 &=  -\frac{2}{2-\xi};
\end{align}
and for the $f$-sector
\begin{align}\nonumber
  \ovl\rho_1 &=  \frac{1}{2}\beta_1 \Omega^{-3} (2+\xi), \mbox{ } &\ovl p_1 &=  -\beta_1
    \Omega^{-3}, \mbox{ } &\ovl\omega_1 &=  -\frac{2}{2+\xi};\\\nonumber
  \ovl\rho_2 &=  \beta_2 \Omega^{-2} (3+\xi), \mbox{ } &\ovl p_2 &= - 3\beta_2
    \Omega^{-2}, \mbox{ } &\ovl\omega_2 &= -\frac{3}{3+\xi};\\
  \ovl\rho_3 &=  \frac{1}{2}\beta_3 \Omega^{-1} (2+\xi), \mbox{ } &\ovl p_3 &=  -\beta_3
    \Omega^{-1}, \mbox{ } &\ovl\omega_3 &= -\frac{2}{2+\xi};
\end{align}
Both groups of equations of state parameters have been plotted in figure
\ref{S:Kerr:fig:plotW-z}. It is obvious that the $\omega$ will diverge for some
value of $\xi$. Nevertheless, this is not a physical problem since the
divergences do not appear in $\rho$ nor $p$. In fact, they are exactly canceled
when multiplied by $\rho$ to obtain $p$. One can note that $\ovl \omega_1 =
\ovl \omega_3$, although their densities and pressures have very different
behaviors due to the different dependence in $\Omega$.

On the other hand, as in section \ref{S:Gordon}, it should be pointed out that
the energy densities follow the relations $\rho_n\propto \left(
\Omega/\Omega_0\right)^{n}$ and $\ovl \rho_n \propto \left(
\Omega/\Omega_0\right)^{n-4}$. In this case, however, the only value of $\zeta$
that gives the same equation of state parameter in both sectors is $\zeta=0$,
which correspond to the conformal transformation. We can relate the equation of
state parameter of each subcomponent in the different sectors through
\begin{equation}
  \ovl\omega_1 = -\frac{\omega_1}{4\omega_1 + 3}, ~~ \ovl\omega_2 =
  -\frac{\omega_2}{2\omega_2 + 1}, ~~ \ovl\omega_3 =
  -\frac{\omega_3}{2\omega_3 + 1}.
\end{equation}

\section{Fast-moving bimetric fluid: Bianchi I solutions}\label{S:Bianchi}
In section \ref{S:FLRW} we considered a FLRW spacetime, which was compatible
with slow moving fluids. In the present case, however, we have a bimetric null
fluid, which in principle  requires an exact treatment of the problem. Thus, we
consider a Bianchi I metric that is compatible with this solution
\cite{BeltranJimenez:2007rsj}
\begin{equation} 
  ds_g^2 =
  -a^2_\parallel d\eta^2 + a^2_\parallel dx^2 + a^2_\perp dy^2 + a^2_\perp dz^2,
  \label{S:Bianchi:eq:g}
\end{equation}
which continues being homogeneous. Here, we have assumed that the 
fluids motion takes place along the x-axis, with $a_\parallel$ and $a_\perp$ the
scale factors parallel and perpendicular to the direction of motion,
respectively. We have also chosen the gauge $\vec S = \vec 0$. We consider the
null vector being $l^\mu = a_\parallel^{-1} (1, 1, 0,0)$ without loss of
generality, as any global factor can be absorbed into $\xi$. As we focus our
attention on solutions satisfying the Kerr-Schild ansatz (\ref{KS}), the metric
of the other gravitational $f$-sector is
\begin{equation}
    ds_f^2 = \Omega^2 [ -a_\parallel^2 (1 - \xi) d\eta^2 -2 a_\parallel^2
      \xi d\eta dx 
    + a_\parallel^2 (1 + \xi)dx^2+ a^2_\perp dy^2 + a^2_\perp dz^2 ].
\label{S:Bianchi:eq:f}
\end{equation}
In this case, the non-vanishing components of the total stress-energy tensor
are
\begin{align}
  T^0_{~0} =& -\sum_\alpha (\rho_{(\alpha)} + p_{(\alpha)})\gamma_\alpha^2 - \rho_N,\\
  T^1_{~0} =& -\sum_\alpha (\rho_{(\alpha)} + p_{(\alpha)})\gamma_{\alpha}^2 v_{(\alpha)} - (\rho_N
  + p_N),\\
  T^0_{~1} =& \sum_\alpha (\rho_{(\alpha)} + p_{(\alpha)})\gamma_\alpha^2 v_{(\alpha)} + (\rho_N
  + p_N),\\
  T^1_{~1} =& \sum_\alpha (\rho_{(\alpha)} + p_{(\alpha)})\gamma_\alpha^2 v^2_{(\alpha)} +
  p_{(\alpha)} + (\rho_N + 2p_N),\\
  T^i_{~i} =& \sum_\alpha p_{(\alpha)} + p_N, ~~ i=2,3,
\end{align}
where the sum goes only through the radiation and dark and baryonic matter
fluids, since we will be  identifying the null bimetric fluid with dark energy.
The $\alpha$-fluid velocity is given by $V_{(\alpha)}^\mu = a_\parallel^{-1}
\gamma_\alpha (1,v_{(\alpha)},0,0)$ with $\gamma_\alpha^2 =
(1-v_{(\alpha)}^2)^{-1}$ from the normalization $g_{\mu\nu}V_{(\alpha)}^\mu
V_{(\alpha)}^\nu = -1$. In this case, we consider all orders in velocities
since we are considering the exact solution.  In addition, as before, the
stress-energy tensor for the material components of the  $f$-sector, $\ovl
T\mmn$, is also of the perfect fluid form.

As the fluids are non-interacting, we can calculate their conservation
equations separately. The exact conservation equations for the null fluid, as
found in reference \cite{BeltranJimenez:2007rsj}, are
\begin{align}
  p_N' =& 0,\\
  (\rho_N+p_N)' + 2(\cH_\parallel +\cH_\perp)(\rho_N + p_N) =& 0,
\end{align}
where we have defined $\cH_\parallel = a'_\parallel/a_\parallel$ and
$\cH_\perp = a'_\perp/a_\perp$, with $'\equiv \partial_\eta$, and whose
solutions are
\begin{eqnarray}
  p_N &=& p_{N0},
  \label{S:Kerr:eq:null-fluid-pressure}\\
  \rho_{N}&=& (\rho_{N0} + p_{N0}) (a_\parallel a_\perp)^{-2} - p_{N0},
  \label{S:Kerr:eq:null-fluid-density}
\end{eqnarray}
and where we have taken, for the sake of simplicity, $a_{\parallel0} =
a_{\perp0} = 1$. Therefore, the equation of state parameter for the null
fluid is
\begin{equation}
  \omega_N = \frac{p_{N0}}{(\rho_{N0} + p_{N0}) (a_\parallel a_\perp)^{-2} - p_{N0}}.
\end{equation}
It can be noted that the null fluid can be split into two contributions: one
that behaves as radiation and other behaving as a cosmological constant. In
contrast, although at late times its equation of state parameter is in
agreement with its energy density behavior ($\omega_N \rightarrow -1$ and
$\rho_N \rightarrow -p_{N0} = \cst$), this is not the case at early times. In
fact, while its density behaves as a radiation fluid, $\rho_N \sim (a_\parallel
a_\perp)^{-2}$, its equation of state parameter approximates to that of matter
($\omega_N \rightarrow 0$). This result comes from the fact that this time the
bimetric fluid is moving at the highest speed.

From equation (\ref{S:Kerr:eq:preassure}) and
(\ref{S:Kerr:eq:null-fluid-pressure}), one can obtain
\begin{equation}
  \Omega(3\beta_1 + 3\Omega \beta_2 + \beta_3 \Omega^2) = - p_{N0},
  \label{S:Bianchi:eq:W}
\end{equation}
where it has been taken into account that null fluid pressure is constant in
time by equation (\ref{S:Kerr:eq:null-fluid-pressure}). This equation implies
that $\Omega$ is constant in time and fixed once $p_{N0}$ is. On the other
hand, from the other equations (\ref{S:Kerr:eq:density+preassure}) and
(\ref{S:Kerr:eq:null-fluid-density}), one gets
\begin{equation}
  \xi = \frac{2 (\rho_{N0} + p_{N0})}{\Omega(\beta_1 + 2\beta_2 \Omega + \beta_3
  \Omega^2)} a_\parallel^{-2} a_{\perp}^{-2} 
  \label{S:Bianchi:eq:xi}
\end{equation}
or equivalently, $\xi = \xi_0 a_\parallel^{-2} a_{\perp}^{-2}$ with
$\xi_0$ fixed once $\rho_{N0}$ and $p_{N0}$ are. This result highly constrains the
possible scenarios. In fact, in massive gravity the reference metric
$f_{\mu\nu}$ must be given by equation (\ref{S:Bianchi:eq:f}) with a constant
$\Omega$, satisfying (\ref{S:Bianchi:eq:W}), and $\xi$ evolving as expressed in
equation (\ref{S:Bianchi:eq:xi}). In bigravity, the material component of the
$f$-sector must be of the form that, through the modified Einstein equations
(\ref{S:notes:eq:Einstein-f}), fixes $\Omega$ and $\xi$ of such form.
Therefore, in this case there is no remaining freedom and the system constants
$\beta_n$ will not qualitatively change the dynamics. 

\begin{figure}[htb]
  \centering
  \includegraphics[width=0.6\textwidth]{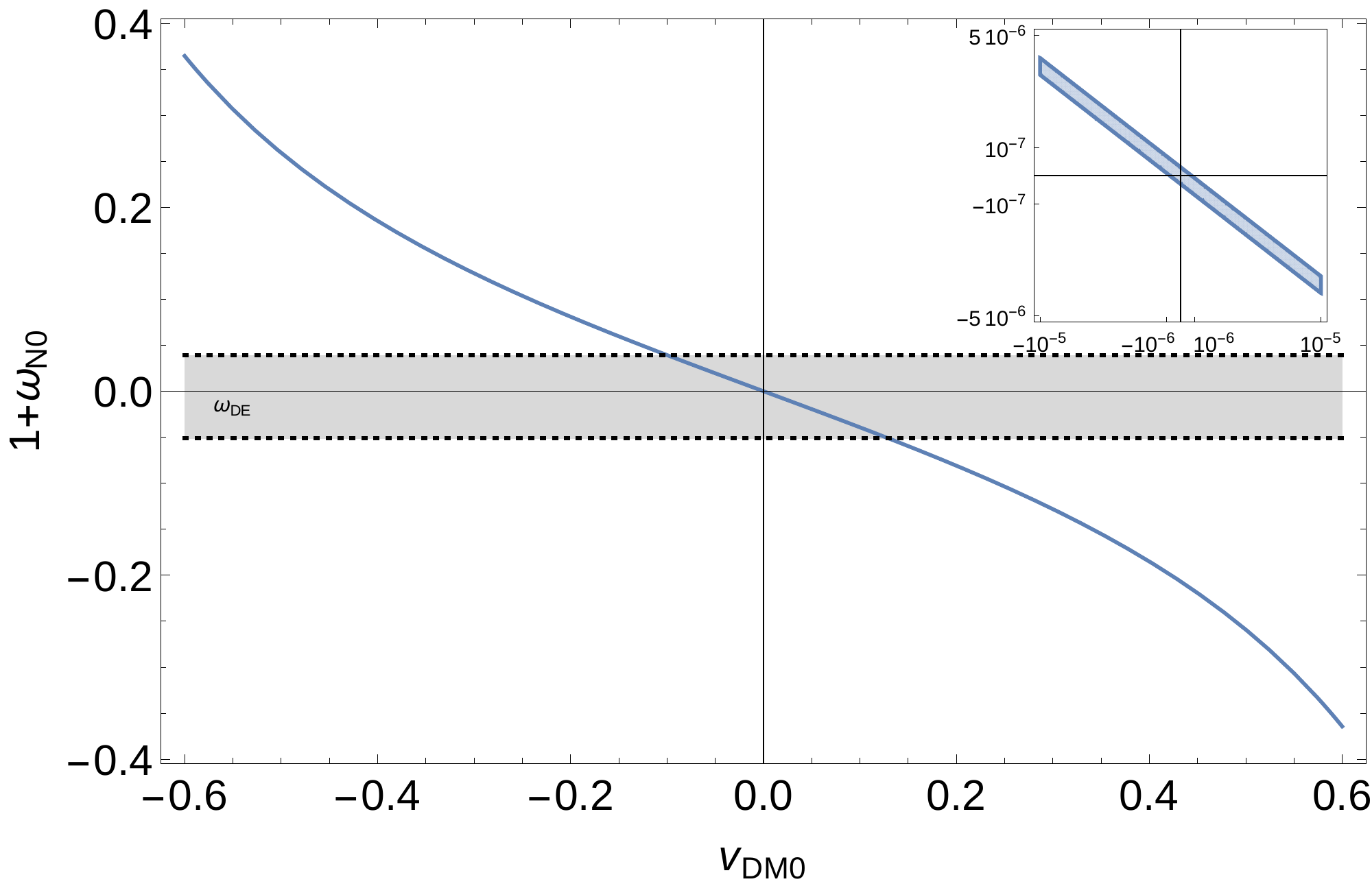}
  \caption{Allowed values of present dark matter velocity related to the
  present value of the equation of state parameter of the null bimetric fluid.
We have supposed $\Omega_N = \Omega_{DE}$. The dashed line is the value of
$\omega_{DE}$ and the dotted lines show its one-sigma region. These data
have been taken from Planck 2015 \cite{Ade:2015xua}: $\omega_{DE} = -1.006 \pm 0.045$.}
\label{S:Bianchi:fig:velocity-general}
\end{figure}

Let us consider the limits imposed by the CMB dipole in this case.  Thus,
imposing  our gauge choice $\vec S = \vec 0$ \cite{BeltranJimenez:2007rsj}
\begin{equation}
  \sum_\alpha (\rho_{(\alpha)} + p_{(\alpha)})\gamma_\alpha^2 v_{(\alpha)} + (\rho_N
  + p_N)=0.
  \label{SNull}
\end{equation}
Nevertheless, as we showed in section \ref{S:FLRW:SS:CMB}, the observed value
of the CMB dipole anisotropy imposes very strong constrains over the possible
values of baryonic matter and radiation velocities, being $|v_{B0}| < 1.1
\times 10^{-6}$ and $|v_R| < 1.23 \times 10^{-3}$, respectively.  In figure
\ref{S:Bianchi:fig:velocity-general}, $\omega_{N0}$ and $v_{DM0}$ are depicted
under the assumption $\Omega_N = \Omega_{DE}$, where the subscript $DE$ denotes dark energy with constant equation of state parameter $\omega_{DE}$. In addition, the value of
$\omega_{DE}$ has also been plotted together with the corresponding one-sigma
region as measured by Planck satellite.  Notice that the dipole constrains the
parameters to lie in a narrow strip along the line shown in figure
\ref{S:Bianchi:fig:velocity-general}, (this is shown in more detail in the
small panel within the figure), thus the allowed range for dark matter
velocity, given a $\omega_{N0}$, is of order $10^{-8}$. In addition, there are
two important conclusions that can be obtained from (\ref{SNull}). On the one
hand, if dark matter moves in the same direction as the other fluids and
$\omega_{N0}<-1$, the bimetric null fluid energy density will be negative while
$p_N > \rho_{N0} (1+\omega_{N0}) a^{-4}$. The reason is that, since its
velocity is completely fixed, density is the only variable that can change sign
to preserve momentum conservation.  On the other hand, as we can see in the
figure, within the one-sigma region of $\omega_{DE}$  the dark matter velocity
could reach non-negligible velocities $-0.13 < v_{DM0} < 0.10$, which could be
interesting within the  bulk flows problem.

Nevertheless, let us study in further detail the case where the cosmic fluids
move slowly and dark energy is given by the null bimetric fluid.  Considering
the Bianchi I metric, the conservation equations to first order in velocity are
as follows:
\begin{itemize}
  \item Energy conservation
    \begin{equation}
      \rho'_{(\alpha)} + (\rho_{(\alpha)} + p_{(\alpha)}) (\cH_\parallel + 2 \cH_\perp) = 0.
      \label{S:Kerr:eq:Energy-conservation-low}
    \end{equation}
  \item Momentum conservation
    \begin{equation}
      \partial_\eta \left[ a_\parallel^2 a_\perp^2 (\rho_{(\alpha)} + p_{(\alpha)})
      v_{(\alpha)} \right] = 0.
      \label{S:Kerr:eq:Momentum-conservation-low}
    \end{equation}
\end{itemize}
Therefore, for low velocities we get
\begin{equation}
  v_{(\alpha)} = v_{\alpha0}\, a_\parallel^{-2} a_\perp^{-2} \frac{1 + \omega_{\alpha0}}{1
  +\omega_\alpha(a)} \frac{\rho_{\alpha0}}{\rho_{(\alpha)}},
  \label{S:Kerr:eq:velocity-low-general}
\end{equation}
and, if the equation of state parameter is constant ($\omega_\alpha=\cst$), the
solutions are
\begin{align}
  \rho_{(\alpha)} =& \rho_{\alpha0} a_\parallel^{-(1+\omega_\alpha)} a_\perp^{-2(1+\omega_\alpha)}
  \label{S:Kerr:eq:density-low},\\
  v_{(\alpha)} =& v_{\alpha_0} a_\parallel^{\omega_\alpha-1} a_\perp^{2\omega_\alpha}.
  \label{S:Kerr:eq:velocity-low}
\end{align}
These equation reduce to those found for a FLRW metric in section \ref{S:FLRW}
(equations (\ref{S:FLRW:eq:density-w-cte}) for density and equation
(\ref{S:FLRW:eq:velocity-w-cte}) for velocity) for past times, since the
null-fluid must have been subdominant until now. Also, as a consequence, the
anisotropy in the past must have been much lower than nowadays. Then, if we
neglected the present anisotropy, it is safer to neglect it at earlier times.
Only in the future the anisotropy may be important enough to be taken into
account.

We consider now the interesting case with dark matter velocity negligible at
present as a consequence of having decoupled from the radiation-baryonic fluid
at early times. In this case we have, $v_{DM0} \simeq 0$ and $|v_R| < 1.23
\times 10^{-3}$, $|v_{B0}| < 1.1 \times 10^{-6}$, from the CMB dipole
observations. Then, the gauge choice $\vec S = \vec 0$ yields
\begin{equation}
  1+\omega_{N0} = - \frac{1}{\Omega_N} \left(\frac{4}{3} \Omega_R v_R + \Omega_{B}
v_{B0}\right), 
\end{equation}
which, choosing the present null fluid energy density to be that of dark energy
in the Standard Model ($\Lambda$CDM), $\Omega_N = \Omega_{DE} \simeq 0.69$
\cite{Ade:2015xua}, one has
\begin{equation}
  |1+\omega_{N0}| <  3.0 \times 10^{-7},
\end{equation}
which is compatible with 2015 Planck data \cite{Ade:2015xua}: $\omega_{DE} =
-1.006 \pm 0.045$.

\begin{figure}[htb]
  \centering
  \begin{subfigure}[b]{0.49\textwidth}
    \includegraphics[width=\textwidth]{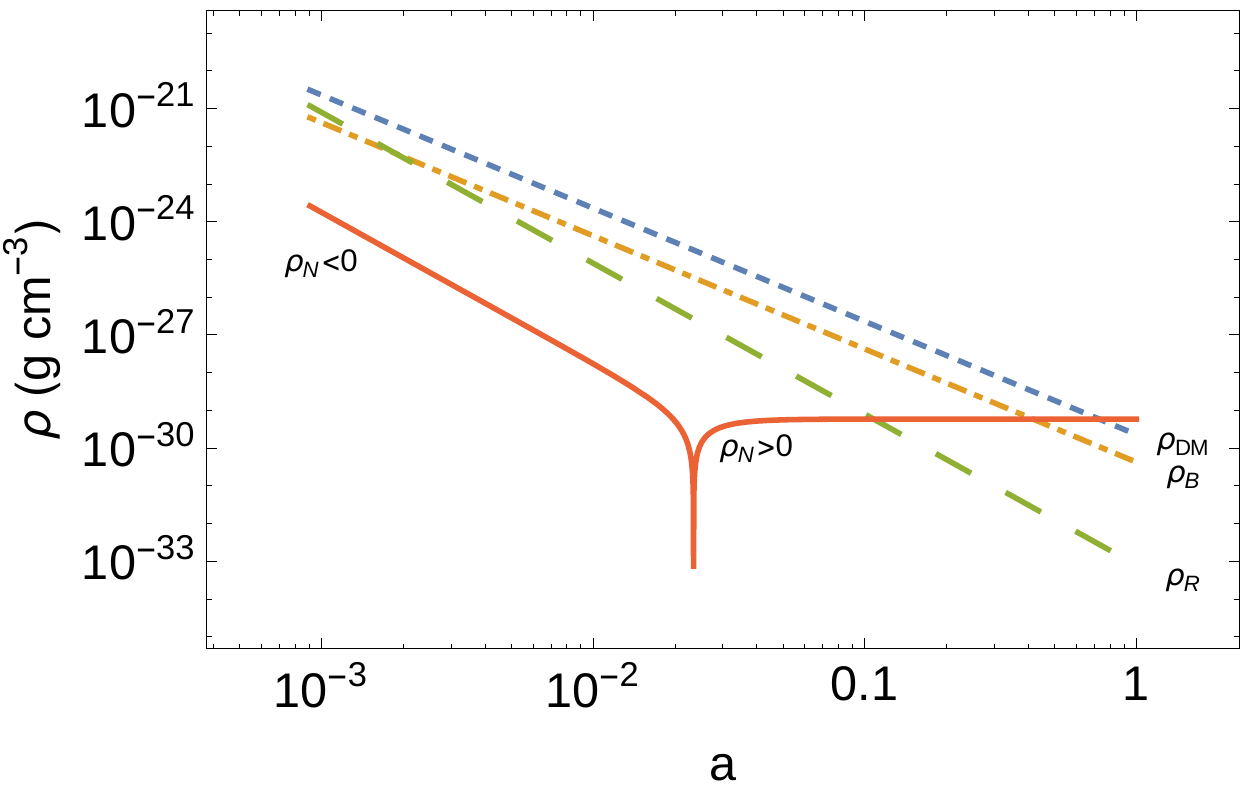}
  \end{subfigure}
  \begin{subfigure}[b]{0.49\textwidth}
    \includegraphics[width=\textwidth]{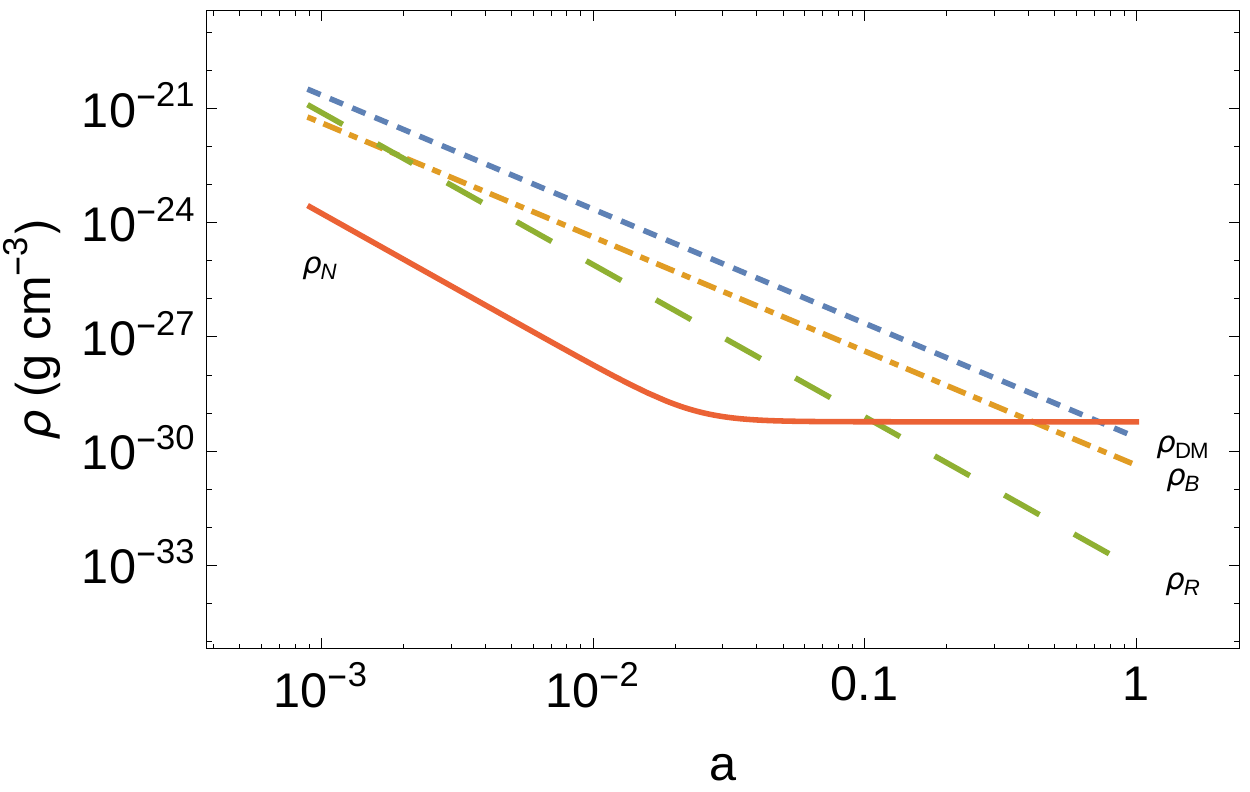}
  \end{subfigure}
  \caption{Energy densities in a universe with radiation, dark and baryonic matter and
    a dark energy null-fluid. As the null-fluid has been subdominant until now,
    we can neglect the anisotropy for past times. The present values
    have been taken to be compatible with CMB dipole, and we have not
    plotted future times when the anisotropy will presumably increase. In the right figure the case of radiation and barionic matter moving in the same direction as dark energy ($v_R, v_B >0$) has been depicted; in this case the energy energy of the null fluid changes sign at $a=0.023$. In the right figure, which corresponds to $v_R, v_B <0$, the null fluid's energy density is always positive.}
  \label{S:Kerr:SS:CMB:fig:CMB}
\end{figure}

\begin{figure}[htb]
  \centering
  \begin{subfigure}[b]{0.49\textwidth}
    \includegraphics[width=\textwidth]{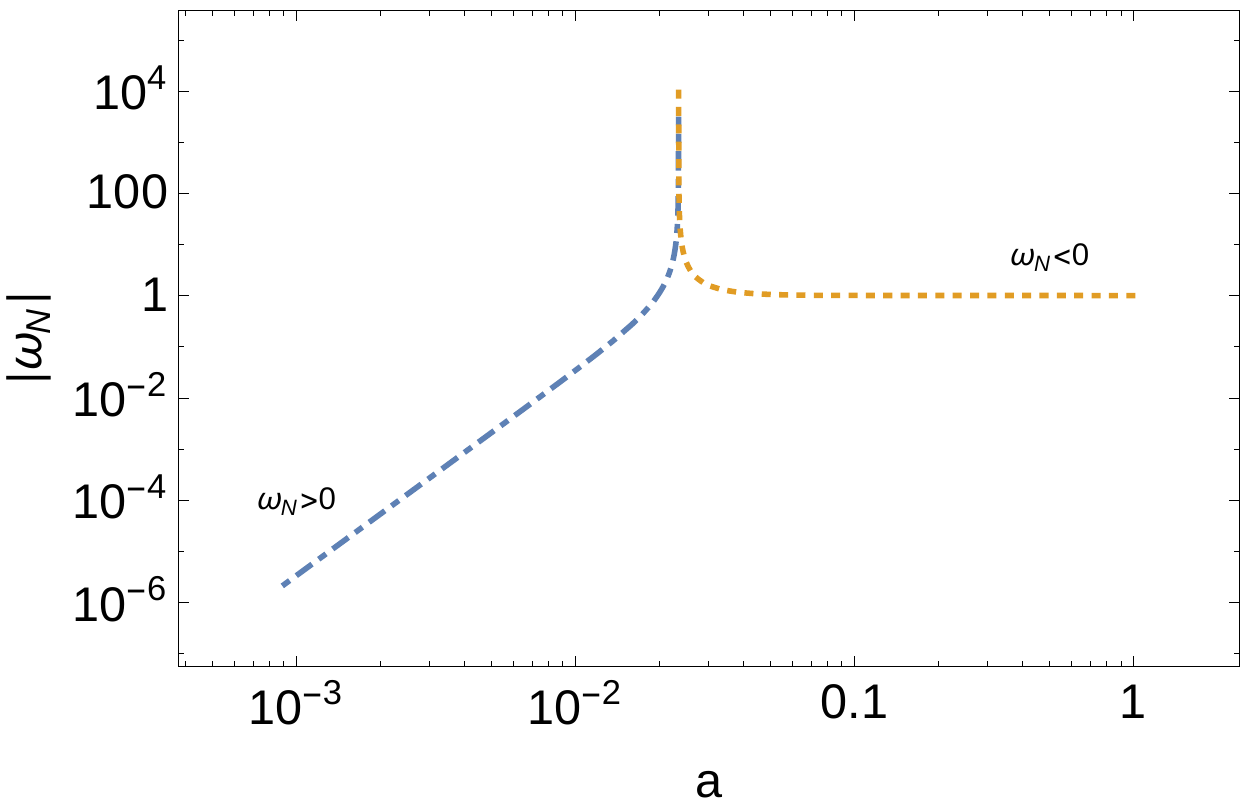}
  \end{subfigure}
  \begin{subfigure}[b]{0.49\textwidth}
    \includegraphics[width=\textwidth]{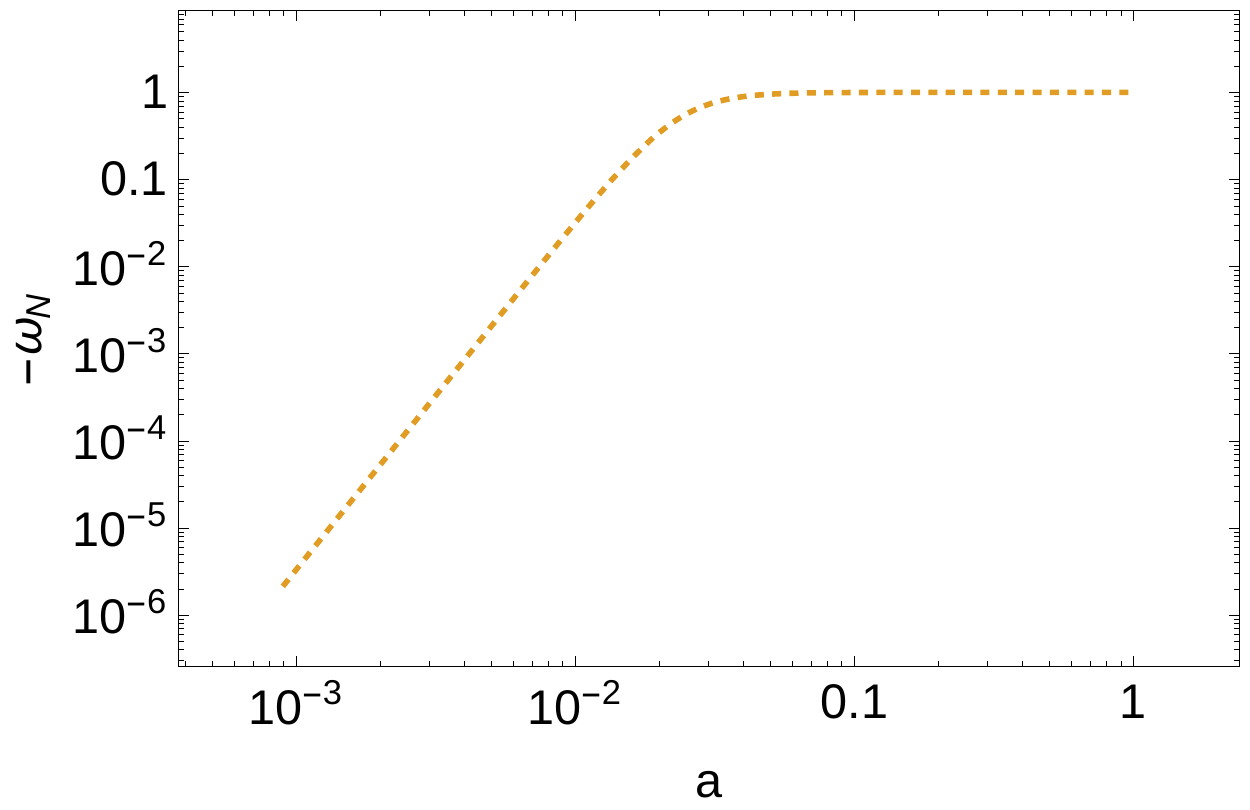}
  \end{subfigure}
  \caption{Null fluid equation of state parameter with present values fixed by
    CMB dipole: $\Omega_{N0} = \Omega_{DE}$ and $(1+\omega_{N0}) =
    \pm 3.0 \times 10^{-7}$. In the right figure, the positive sign has been
    taken in the previous equation whereas in the left one we have considered
    the negative sign. The change of sign and the peak exhibited on the later
    is caused by the change of sign of $\rho_N$ at $a = 0.023$.}
  \label{S:Kerr:SS:CMB:fig:w}
\end{figure}

\begin{figure}[htb]
  \centering
  \includegraphics[width=0.6\textwidth]{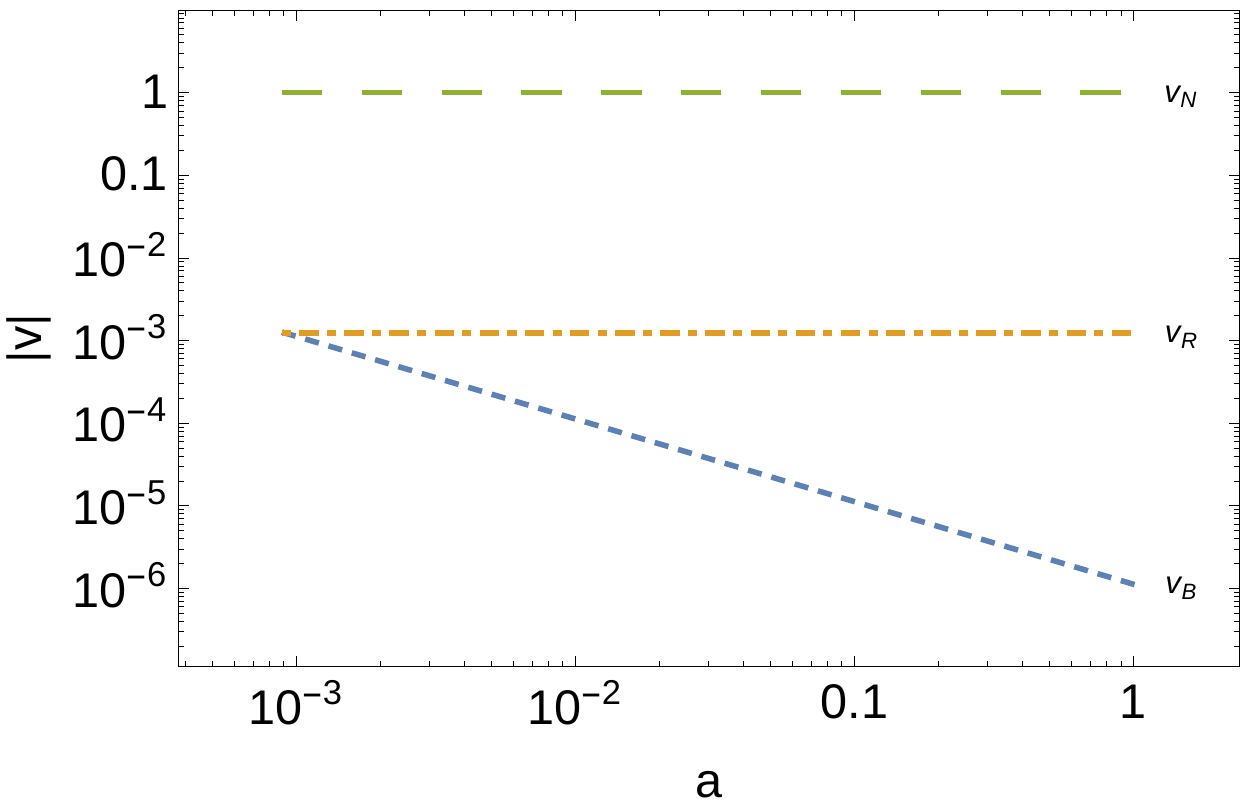}
  \caption{Velocities for a universe filled with radiation, dark and baryonic
  matter and a dark energy null-fluid. The present values have been taken to
be compatible with CMB dipole, in the case with dark matter at rest.}
    \label{S:Kerr:SS:CMB:fig:CMB-v}
\end{figure}

In figure \ref{S:Kerr:SS:CMB:fig:CMB} the fluid energy densities have been
plotted. One can see that, depending on the relative direction of the motion of
the bimetric fluid and the radiation and baryonic matter, one can have positive
or negative energy densities for the bimetric fluid, since being null, the
bimetric fluid velocity is fixed. In fact, when the other two fluids move in
the same sense as the null fluid, its energy density has to be negative for
times earlier than $a=0.023$ in order to cancel the momentum of the other two
fluids. One can compare the density behavior with the equation of state
parameter plotted in figure \ref{S:Kerr:SS:CMB:fig:w}. In both cases, the
figures for parallel moving fluids show a peak as a result of the change of
sign of the density. It must also be noted that the null dark fluid has been
subdominant until present, when it starts to dominate. Finally, fluids
velocities have been plotted in figure \ref{S:Kerr:SS:CMB:fig:CMB-v}. 

\section{Conclusions}\label{S:Conclusion}
In this work we have studied the cosmological consequences of massive gravity
and bigravity solutions whose metrics are not necessarily diagonal in the same
patch of coordinates, as assumed in previous literature.  Focusing our
attention on solutions in which both causal structures have a particular
relation, we have explored situations in which the interaction between both
metrics is described by an effective bimetric fluid with an apparent velocity
with respect to the comoving coordinates.

In the first place, we have considered solutions with both metrics related
through the generalized Gordon ansatz, which implies that the light cones of one of the metrics are strictly inside those of the
other metric. As the conserved stress-energy
tensor for the effective bimetric fluid of these solutions is of the perfect
fluid form, with a four-velocity given by the timelike vector of the ansatz,
and there are three free parameters of the theory apart from those leading to
cosmological constant contributions, we have been able to decompose the
bimetric fluid in three perfect fluid components with the same velocity.
Although these components are not separately conserved in general, their
effective equation of state parameters depend only on one function of the
ansatz, allowing to determine the behavior of the effective fluids that could
coexist in the same bimetric solution.

Considering these bimetric solutions in a cosmological scenario, we have been
able to describe a universe that satisfies the cosmological principle in the
limit of a slow-moving perfect bimetric fluid, that is the case in which the
timelike vector of the Gordon ansatz does not depart significantly from the
comoving vector. As we have chosen to work in the cosmic center of mass frame,
the material content of our gravitational sector has to be in motion. In this
scenario, the energies of the three bimetric  components are individually
conserved but their momenta are not. Restricting attention to theories in which
the effective bimetric fluid only has one of these components, we have explored
in detail the cases where the bimetric fluid can be identified with dark
radiation, dark matter and a fluid behaving as dark matter at early times and
as a cosmological constant at late times. Each case corresponds to consider a
particular theory of massive gravity, with fixed reference metric, or a
hypothetical material content hidden in the gravitational sector that we are
not inhabiting compatible with that solution in bigravity.  Moreover, we have
been able to constrain the possible values of the velocities of the fluids
taking into account the data available about the density parameters and the
magnitude of the CMB dipole. On the other hand, we have also considered an
effective fluid with two components, due to two non-vanishing parameters in the
interaction Lagrangian, obtaining a new purely bimetric phenomenon consisting
of a transition from an accelerating to a decelerating fluid velocity.

In the second place, we have studied the effects of a null bimetric fluid,
which can be obtained by restricting attention to solutions that satisfy the
Kerr-Shild ansatz. The null vector describing the motion of the effective fluid
in this case is just the only common null vector of both metrics. Therefore,
this is the opposite limit to the slowly moving bimetric fluid that we had
previously studied. In this case, one necessarily has to consider an exact
solution to describe our Universe, which can be a Bianchi I spacetime. Although
the bimetric fluid can also be decomposed in three components, such
decomposition is not so useful as in the previous case since the free functions
appearing in the ansatz (and, therefore, in the fluid) are completely fixed
once the conservation equation for the null fluid is considered in this
scenario. Moreover, such fixing also implies that the scenario is only
compatible with a group of massive gravity theories, with reference metrics set
accordingly, or that a particular material content has to be present in the
other gravitational sector if the cosmological solution is due to a bigravity
theory. As in the slow-moving fluids case, we have used the available data to
constrain the fluids velocities. In this framework it is possible, at least in
principle, to obtain scenarios beyond the slowly moving dark matter case. On
the other hand, we have also constrained the null fluid equation of state
parameter identifying this fluid with dark energy and taking into account that
the rest of cosmic fluids moves slowly.

\begin{acknowledgments}
The authors acknowledge Matt Visser for useful discussions. This work has been
supported by the Spanish MICINNs Consolider-Ingenio 2010 Programme under grant
MultiDark CSD2009-00064 and MINECO grant FIS2014-52837-P. PMM also acknowledges
financial support from the Spanish Ministry of Economy and Competitiviness
through the postdoctoral training contract FPDI-2013-16161. CGG is supported by
a UCM-Manuel \'Alvarez L\'opez grant.  
\end{acknowledgments}

\end{document}